\documentclass[structabstract]{aa}  
\usepackage{fancyvrb}
\usepackage{multirow}
\usepackage{graphicx}
\usepackage{color,xcolor}

\definecolor{orange}{HTML}{FF7F00}
\definecolor{dgreen}{HTML}{339933}

\usepackage{txfonts}
\begin{document}
\VerbatimFootnotes
   \title{
   Stripped gas as fuel for newly formed HII regions in the encounter between
   VCC1249 and M49: a unified picture from NGVS and GUViCS.}

   \author{Fabrizio Arrigoni Battaia \inst{1,2}, Giuseppe Gavazzi \inst{2},
          Michele Fumagalli\inst{3},  Alessandro Boselli \inst{4}, Samuel Boissier \inst{4}, 
          Luca Cortese \inst{5}, Sebastien Heinis \inst{4},
	  Laura Ferrarese \inst{6}, Patrick C\^ot\'e \inst{6}, J. Christopher Mihos \inst{10},  
	  Jean Charles Cuillandre \inst{7}, Pierre-Alain Duc \inst{12}, Patrick Durrell \inst{11},
	  Stephen Gwyn \inst{6},
	   Andr\'es Jord\'an \inst{9}, Chengze Liu \inst{13},  Eric Peng \inst{8}, Simona Mei \inst{14,15,16}
	             }
   \institute{\scriptsize
   	Max-Planck-Institut f\"ur Astronomie, K\"onigstuhl 17, D-69117 Heidelberg, Germany \\
	    		 \email{arrigoni@mpia.de} 
	    		 \and
  	Dipartimento di Fisica G. Occhialini, Universit\`a di Milano- Bicocca, Piazza della Scienza 3, 20126 Milano, Italy\\
             \email{giuseppe.gavazzi@mib.infn.it} 
	     \and
	Department of Astronomy and Astrophysics, University of California, 1156 High Street, Santa Cruz, CA 95064, USA\\    
            \email{mfumagalli@ucolick.org}
		\and
	Laboratoire d'Astrophysique de Marseille - LAM, Universit\'e d'Aix-Marseille \& CNRS, UMR7326, 38 rue F. Joliot-Curie, 13388 Marseille Cedex 13, France\\
         \email{alessandro.boselli@oamp.fr; samuel.boissier@oamp.fr; Sebastien.heinis@oamp.fr} 
	\and
	European Southern Observatory, Karl-Schwarzschild Strasse 2, 85748 Garching bei Muenchen, Germany\\
	\email{lcortese@eso.org}
	\and 
	Herzberg Institute of Astrophysics, National Research Council of Canada, Victoria, BC V9E 2E7, Canada\\
	\email{Laura.Ferrarese@nrc-cnrc.gc.ca; patrick.cote@nrc-cnrc.gc.ca: Stephen.Gwyn@nrc-cnrc.gc.ca}
	\and
	CFHT corporation, 65-1238 Mamalahoa Highway, Kamuela, Hawaii 96743, USA\\
        \email{jcc@cfht.hawaii.edu}
	\and
	Kavli Institute for Astronomy and Astrophysics, Peking University, Beijing 100871, China\\
	\email{peng@pku.edu.cn}
	\and
	Departamento de Astronom\'ia y Astrof\'isica, Pontificia Universidad Cat\'olica de Chile, Vicuna Mackenna 4860, 7820436 Macul, Santiago, Chile\\
	\email{ajordan@astro.puc.cl}
	\and
	Department of Astronomy, Case Western Reserve University, 10900 Euclid Ave, Cleveland, OH 44106, USA\\
	\email{mihos@case.edu}
	\and
	Department of Physics \& Astronomy, Youngstown State University, Youngstown, OH 44555, USA\\
	\email{prdurrell@ysu.edu}
	\and
	Laboratoire AIM, CEA/DSM-CNRS-Universit\'e Paris Diderot, Dapnia/Service d'Astrophysique, CEA-Saclay, 91191 Gif-sur-Yvette Cedex, France\\
	\email{paduc@cea.fr}
	\and
	INPAC and Department of Physics, Shanghai Jiao Tong University, 800 Dongchuan Road, Shanghai 200240, China\\
	\email{czliu@pku.edu.cn}
	\and
        GEPI, Observatoire de Paris, Section de Meudon, 5 Place J. Janssen, 92190 Meudon Cedex, France\\ 
	\email{simona.mei@obspm.fr}
	\and
	Universit\'{e} Paris Denis Diderot, 75205 Paris Cedex 13, France
	\and
	California Institute of Technology, Pasadena, CA 91125, USA\\ 
	 }
              \date{Received 26/01/2012; accepted 30/04/2012}

 
  \abstract
   { 
     We study the peculiar interacting galaxy system of VCC1249/M49 located in the core 
     of the Virgo B subcluster. 
     Owing to a recent interaction between the dwarf galaxy VCC1249 and the halo gas of the 
     elliptical galaxy M49, neutral hydrogen has been displaced from 
     the interstellar medium of this dwarf into the Virgo intracluster medium. 
     Observations also reveal multiple compact star-forming regions (\emph{aka} HII regions)
     that are embedded in this HI cloud, with a projected separation up to 13 kpc from 
     VCC1249 in the northwest direction.
   }
   {
     Motivated by recent near-ultraviolet (NUV) imaging from the GALEX Ultraviolet Virgo Cluster Survey (GUViCS) 
     of the VCC1249/M49 system that shows significant 
     ongoing/recent star formation in the compact regions, we aim to constrain the
     origin of these outlying HII regions with a multi-wavelength approach.
   }
   {
       Using deep optical ($u, g, i, z$) imaging from the Next Generation Virgo 
       Cluster Survey (NGVS) and new H$\alpha$ imaging obtained at the 
       San Pedro Martir observatory together with Keck long-slit spectroscopy, we characterize
       the star formation rates, ages, and metallicity of VCC1249 and its 
       outlying compact regions. Moreover, we analyze the color and luminosity profile 
       of the galaxy to investigate its recent interaction with M49.
   }
   {
   Our new observations indicate that VCC1249 underwent a recent 
   interaction with M49 in which both ram-pressure stripping and tidal interaction occured. 
   The joint action of the two mechanisms led to the removal of the HI gas from the interstellar medium of VCC1249, 
   while the gravitational tides triggered the stellar tail and counter-tail of VCC1249.
   Our stellar population synthesis analysis 
   reveals that the star formation in this galaxy  was truncated 
   around 200 Myr ago and that the outlying HII 
   regions were born \emph{in situ} $\approx$ 10 Myr ago out of 
   pre-enriched gas removed from the dwarf galaxy. These observations also reveal 
   that interactions between central and satellite galaxies similar to those between VCC1249/M49 
   may be an effective way of dispersing metals into the halos of massive galaxies.
   }
   {}

   \keywords{Galaxies: clusters: individual: Virgo; Galaxies evolution; Galaxies irregular}

%
\authorrunning{Arrigoni Battaia et al.}
\titlerunning{Stripped gas as fuel for newly formed HII regions 
in the encounter between VCC1249 and M49} 
\maketitle

\section{Introduction}
The role of the environment in shaping the observed properties of galaxies 
(morphology, star formation, color, gas content, etc.) 
during their evolution is the subject of an ongoing debate.
Among the multiple processes that operate 
in dense environments (see the review by Boselli \& Gavazzi 2006),
ram-pressure stripping (Gunn \& Gott 1972)  is often invoked as the principal mechanism 
acting in clusters, especially on dwarf galaxies.
Indeed, due to their shallower gravitational potential, 
the dynamical pressure that originates from 
the fast motion of these galaxies through the hot and dense intra-cluster medium (ICM)
can easily exceed the gravitational binding force in the galaxy
and efficiently remove atomic gas from their interstellar medium (ISM). 
The resulting sudden suppression of the star formation in these stripped galaxies, then, 
leads to their temporary transformation into post star-burst (K+A; Poggianti et al. 2004), 
and subsequently into dE galaxies (Gavazzi et al. 2010). 
Ram-pressure stripping can therefore be the leading mechanism for the 
migration of dwarf galaxies from the blue cloud to the red sequence 
(Boselli et al. 2008a,b).

Stripped neutral hydrogen is frequently observed in the local Universe.
For instance, extended HI tails are observed in the 
Virgo cluster (Chung et al. 2007, 2009) 
associated with late-type galaxies within 1 Mpc projected
distance from M87. Because the stripped gas always points away from the cluster 
center, these tails are a clear signature of the ram-pressure mechanism. Another 
typical feature of a ram-pressure stripped tail is the absence of a stellar counterpart.
Moreover, galaxies that are infalling in rich clusters occasionally show 
stripped gas with associated star formation (Cortese 
et al. 2007), as evident from their ultraviolet (UV) (Smith et al. 2011), H$\alpha$  (Yagi et al. 2010), 
and X-ray emission (Sun et al. 2010).
A remarkable example is the dwarf irregular (dIrr) galaxy 
VCC1217, which shows an extended tail of bright knots and of diffuse emission in the UV light 
(Hester et al. 2010; Fumagalli et al. 2011), which is 
associated to star formation triggered by the interaction with the intergalactic medium.
This interpretation is also consistent with the findings of 
hydrodynamic simulations (Kapferer et al. 2009).

In addition to hydrodynamic processes, tidal interactions  
(e.g. Spitzer \& Baade 1951; Valluri \& Jog 1990; Duc \& Renaud 2011) can efficiently remove
baryons from galaxies, in particular from the outer and less bound regions.
However, because of the high relative velocities among cluster galaxies, these encounters
occur on shorter time scales than in the field,  
resulting in a less severe disturbance of the galaxy stellar and gaseous disks.
Moreover, galaxies in clusters grow through merging and accretion processes,
consistent with hierarchical models and as indicated by the extensive web of low-surface 
brightness filaments and tidal features 
found by Janowiecki et al. (2010) in the intracluster light around luminous 
ellipticals in  Virgo (M49, M87, M86, M89). 
Perhaps the most spectacular example of an interacting system in the Virgo cluster is that 
between NGC4438 and M86,
whose morphology and gas content require that
tidal interaction and ram-pressure stripping are simultaneously occuring (Kenney et al. 2008). 
Similarly, the perturbations observed in NGC4654 and NGC4254 can be modeled with a
close encounter with a nearby companion combined with ram-pressure 
stripping (Vollmer 2003; Vollmer et al. 2005).

Environmental processes may also play a role in the 
formation of ultra compact dwarfs (UCD; Drinkwater et al. 2000), a new  
class of objects in clusters whose stellar masses exceed those observed in 
the brightest globular clusters only by small amounts (Ha{\c s}egan et al. 2005; Mieske et al. 2008).
However, their origin is controversial. One hypothesis is that they derive from stripping 
of dE galaxies in the cluster potential or the potential of its brightest
members (``threshing scenario'' Bekki et al. 2003). Alternatively, 
a merger of young and massive star clusters that formed during an interaction between 
gas-rich galaxies (Kroupa 1998; Fellhauer \& Kroupa 2002) could account for the origin of
UCDs.

Similar arguments apply to the rare class of compact elliptical (cE) galaxies, such as M32, whose 
compact nature has been ascribed to tidal stripping (e.g., Faber 1973; Bekki et al. 2001; 
Huxor et al. 2011).

In this paper, we focus on the interaction between the dIrr galaxy VCC1249 ( \emph{aka} UGC 7636) 
and the massive elliptical M49. The system VCC1249/M49 has been studied previously, 
following the detection of an HI cloud displaced from the galaxy in the direction of 
M49 (Sancisi et al. 1987;  Patterson \& Thuan 1992;  Henning et al. 1993). 
Subsequently, McNamara et al. (1994) found a trail of debris that does not coincide with the 
HI gas and is projected northward from the dwarf galaxy\footnote{ This feature and the 
similarity in shape of the HI cloud and galaxy (see Figure \ref{withHI}) are the two strongest signatures 
for the ram-pressure mechanism highlighted by McNamara et al. (1994).}. Lee et al. (1997, 2000) furthermore
showed the presence of young HII regions embedded in the HI cloud. 
The conclusions of these studies is that both ram-pressure stripping and tidal interaction 
are necessary to explain the morphology and the deficiency of HI in VCC1249.

Our attention was again caught by the VCC1249/M49 system owing to a recent deep NUV GALEX image, 
obtained as part of the GALEX Ultraviolet Virgo Cluster Survey (GUViCS) (Boselli et al. 2011), 
showing an extended UV feature stretching from VCC1249 toward M49, with some
relative maxima in correspondence with the detached HI emission. By combining these new data 
to high-quality optical imaging data from the Next Generation Virgo 
Cluster Survey (NGVS) (Ferrarese et al. 2011) at CFHT and with 
Keck optical spectroscopy of the external HII regions embedded in the HI cloud, 
we are able to conduct the first multiwavelength analysis of the properties of VCC1249 
and of the outlying HII regions. In particular, following a  procedure similar to 
Fumagalli et al. (2011) for VCC1217, we adopt a spectral energy distribution (SED) 
fitting technique to constrain the age of external HII regions and the epoch at 
which the star formation in VCC1249 was suddenly truncated, presumably by the 
interaction with M49.
We stress that the ram-pressure mechanism we are referring to throughout this work 
is not caused by the ICM within the Virgo Cluster, but by the hot and dense halo of M49,
as described in Mayer et al. (2006).

The organization of the present paper is as follows: in $\mathcal{x}$ 2, we discuss the observations and 
data reduction for each band. The properties of VCC1249 and of the external star-forming regions 
are studied in $\mathcal{x}$ 3 and $\mathcal{x}$ 4, respectively. In $\mathcal{x}$ 5 we show the results 
of the SED fitting analysis for both VCC1249 and the HII regions and in $\mathcal{x}$ 6 we summarize our 
conclusions and present a panoramic view of the Virgo's B subcluster updated accordingly.

Throughout this paper we assume a standard cosmology and a distance modulus of 
31.14$\pm$0.05 mag for M49 corresponding to the distance of 16.9$\pm$0.3 Mpc 
(Mei at al. 2007), fully consistent with 16.7$\pm$0.5 Mpc from
Blakeslee et al. (2009) and with 17.0 Mpc from Gavazzi et al. (1999).
Magnitudes are given in the AB system throughout the paper.

\section{Observations and data reduction}
\label{The data}

\begin{figure*}
\centering
\includegraphics[scale=0.96]{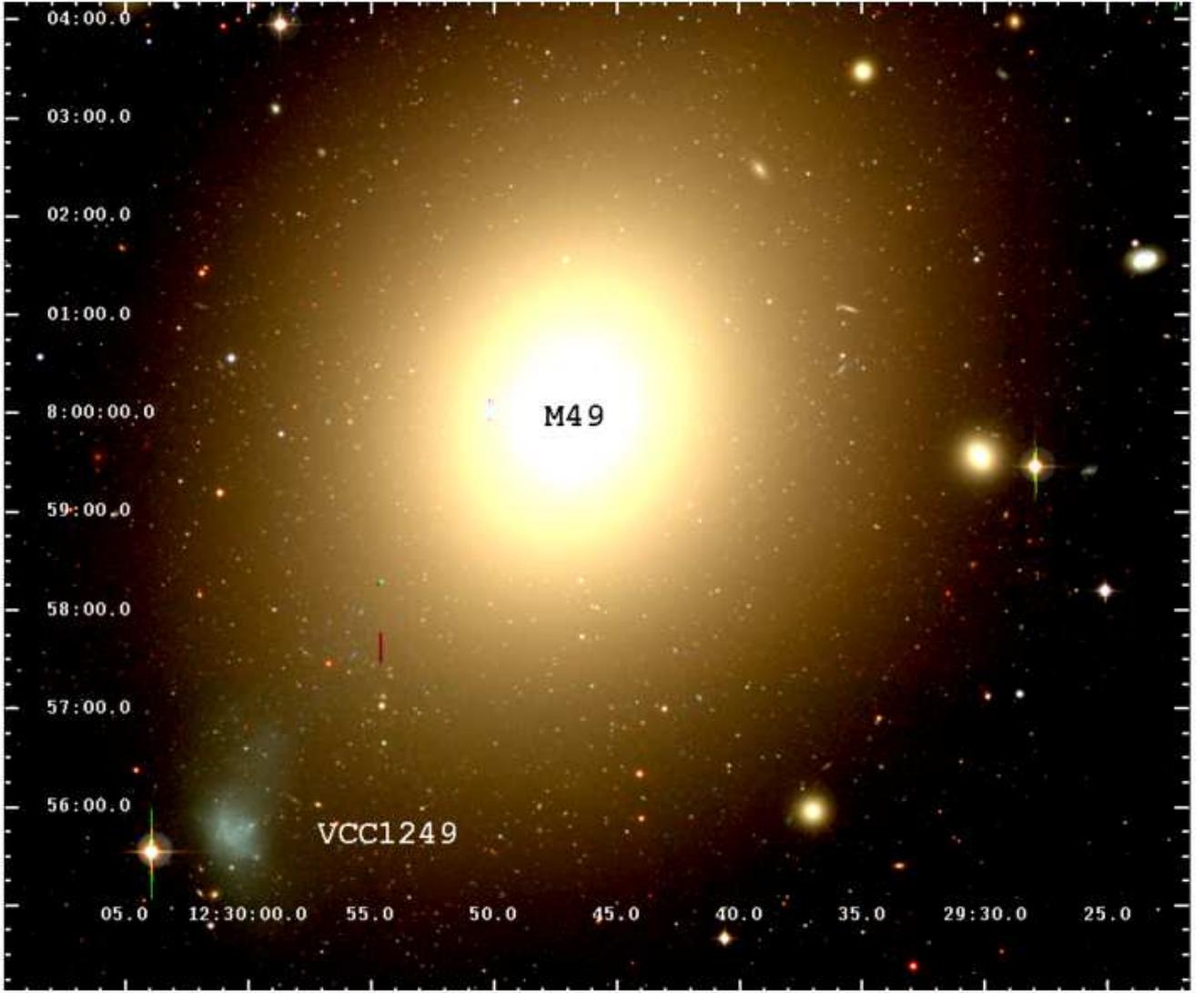} 
\caption{RGB image of VCC1249 (bottom-left) and of M49 (center) obtained by combining the NGVS images 
in the $u, g, z$ filters. The difference in color between the red giant elliptical M49 and the 
blue dIrr VCC1249 is apparent. The faint tidal tail of VCC1249 is also visible in the northwest 
direction.}
\label{Fig1}
\end{figure*}

\begin{table}
\caption{Log of the observations.}
\centering
{\footnotesize \begin{tabular}{lcllc}
\hline
\hline
\multicolumn{5}{c}{Imaging}\\
\hline
Telescope & Filter & Seeing    & Date         & Exp. Time\\
          &        & (arcsec)  & (yyyy-mm-dd) & (sec)\\
\hline	   
CFHT & $u$-NGVS & 0.96  & 2009-02-27    & 6402  \\
CFHT & $g$-NGVS & 0.71  & 2009-02-22    & 3170  \\
CFHT & $i$-NGVS & 0.61  & 2009-02-18    & 2050   \\
CFHT & $z$-NGVS & 0.82  & 2009-02-22    & 4400   \\
\\
GALEX & NUV & 5.3    & 2010-03-19    & 1631 \\
GALEX & FUV & 4.2    & 2004-04-17    & 104  \\
\\
SPM & $H\alpha$ 6603 & 1.5  & 2011-04-25 & 3$\times$600  \\
SPM & $r'$ & 1.5  & 2011-04-25 & 300 \\
\hline
\multicolumn{5}{c}{Spectroscopy}\\
\hline	
Telescope & Instrument & Slit/Fiber    & Date         & Exp. Time\\
          &            & (arcsec)      & (yyyy-mm-dd) & (sec)\\
\hline
Keck  & LRIS  &  1.0  & 2011-01-3 & 2$\times$1200  \\
Keck & LRIS &    1.0  & 2011-01-4  & 900 \\
Lick & KAST &    2.0  & 2011-04-29 & 2$\times$1800\\
SDSS & SDSS &    3.0  & DR7 & 2040  \\
\hline
\end{tabular}
}
\label{Table1}
\end{table}

Since 2009 we have collected new deep 
observations of the system VCC1249/M49, covering a wide
stretch of the electromagnetic spectrum, including UV (NUV), optical imaging ($u, g, i, z$ + $H\alpha$),
and optical spectroscopy. A summary of these observations is presented in Table \ref{Table1}.
In this section, we present a brief description of these observations and of the data reduction.

Images and spectra were analyzed
using the STSDAS and GALPHOT\footnote{Developed for IRAF - 
STSDAS mainly by W. Freudling, J. Salzer, and
M.P. Haynes (see Haynes et al. 1999) and adapted by L. Cortese and S. Zibetti to handle H$\alpha$ data.} routines in the
IRAF\footnote{IRAF is the Image Analysis and
Reduction Facility made available to the astronomical community by the
National Optical Astronomy Observatories, which are operated by AURA,
Inc., under contract with the U.S. National Science Foundation. STSDAS
is distributed by the Space Telescope Science Institute, which is
operated by the Association of Universities for Research in Astronomy
(AURA), Inc., under NASA contract NAS 5--26555.} 
and FUNTOOLS\footnote{Developed for DS9 by the High Energy Astrophysics Division of SAO.} packages.

\begin{figure*}
  \centering
\begin{tabular}{cc}
  \includegraphics[scale=0.48]{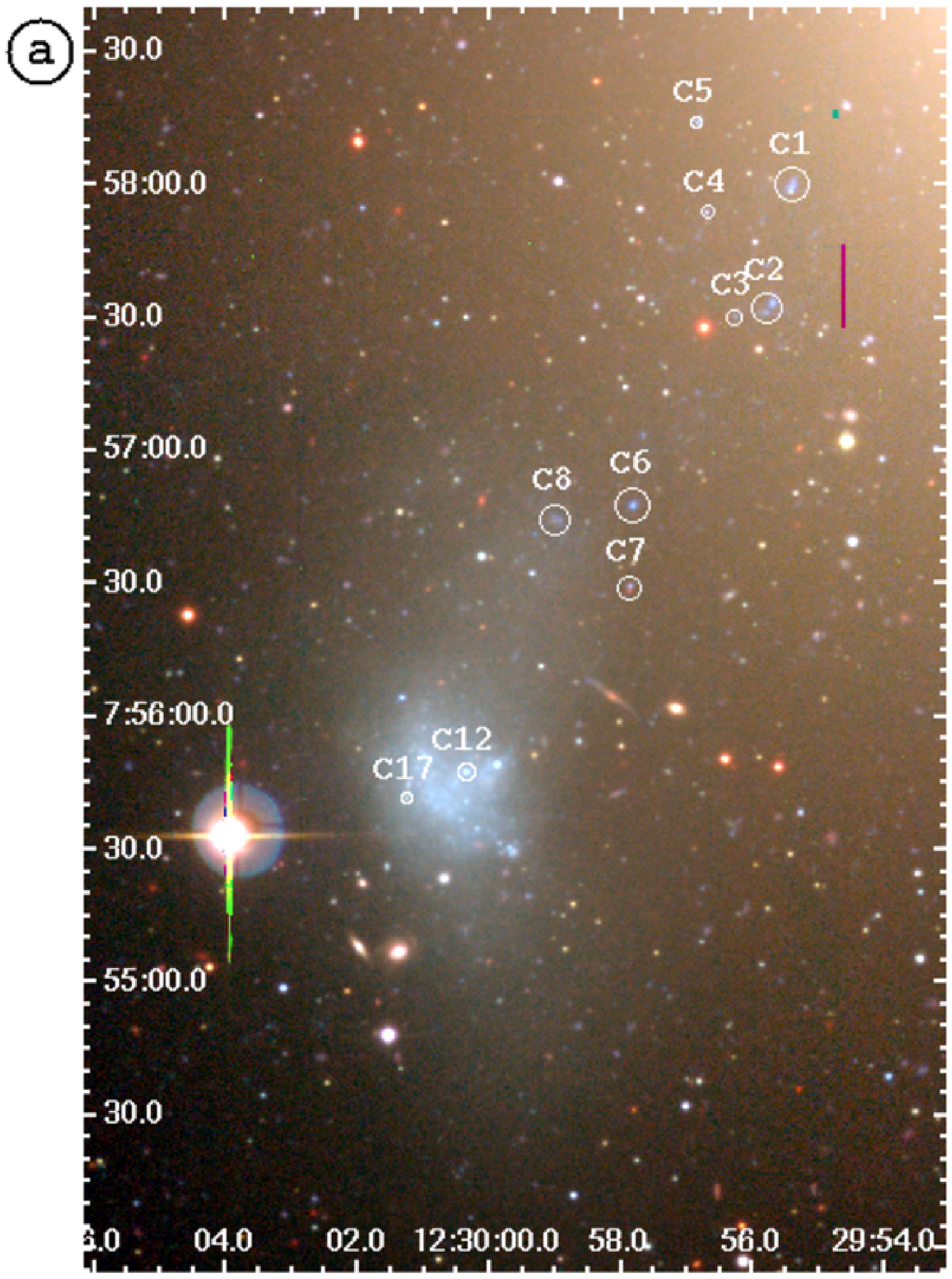}& 
  \includegraphics[scale=0.48]{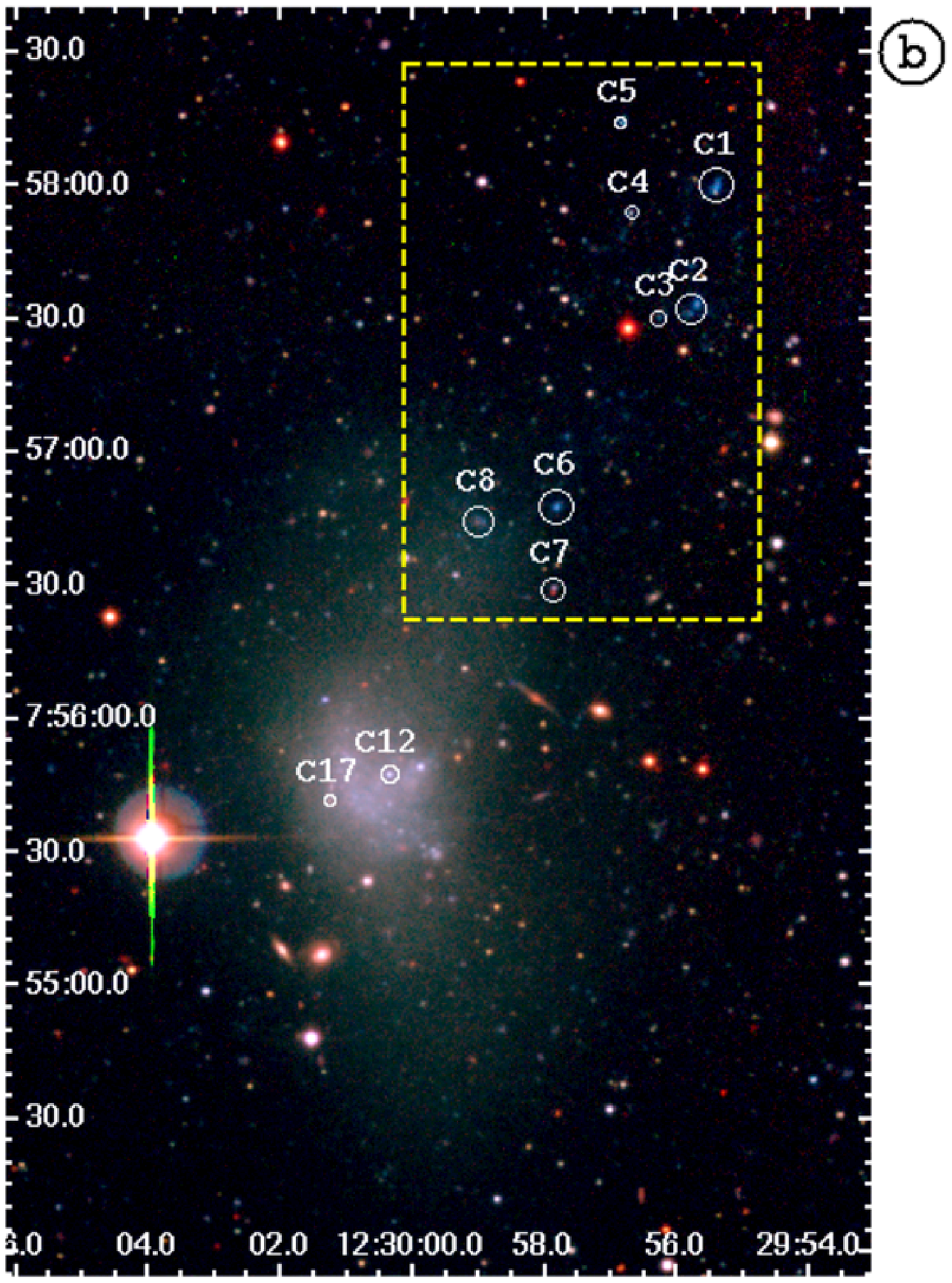}\\
  \includegraphics[scale=0.48]{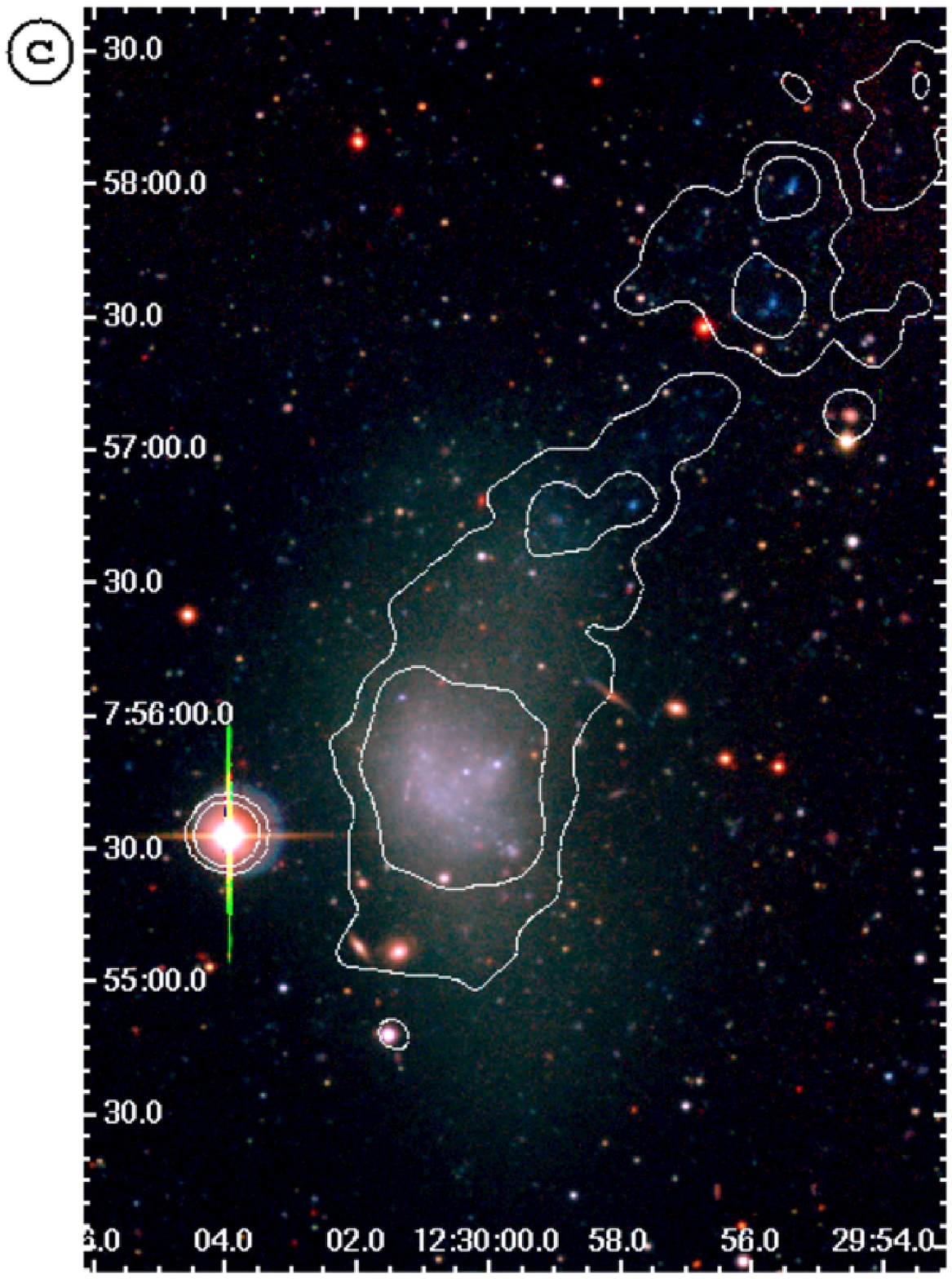}& 
  \includegraphics[scale=0.479]{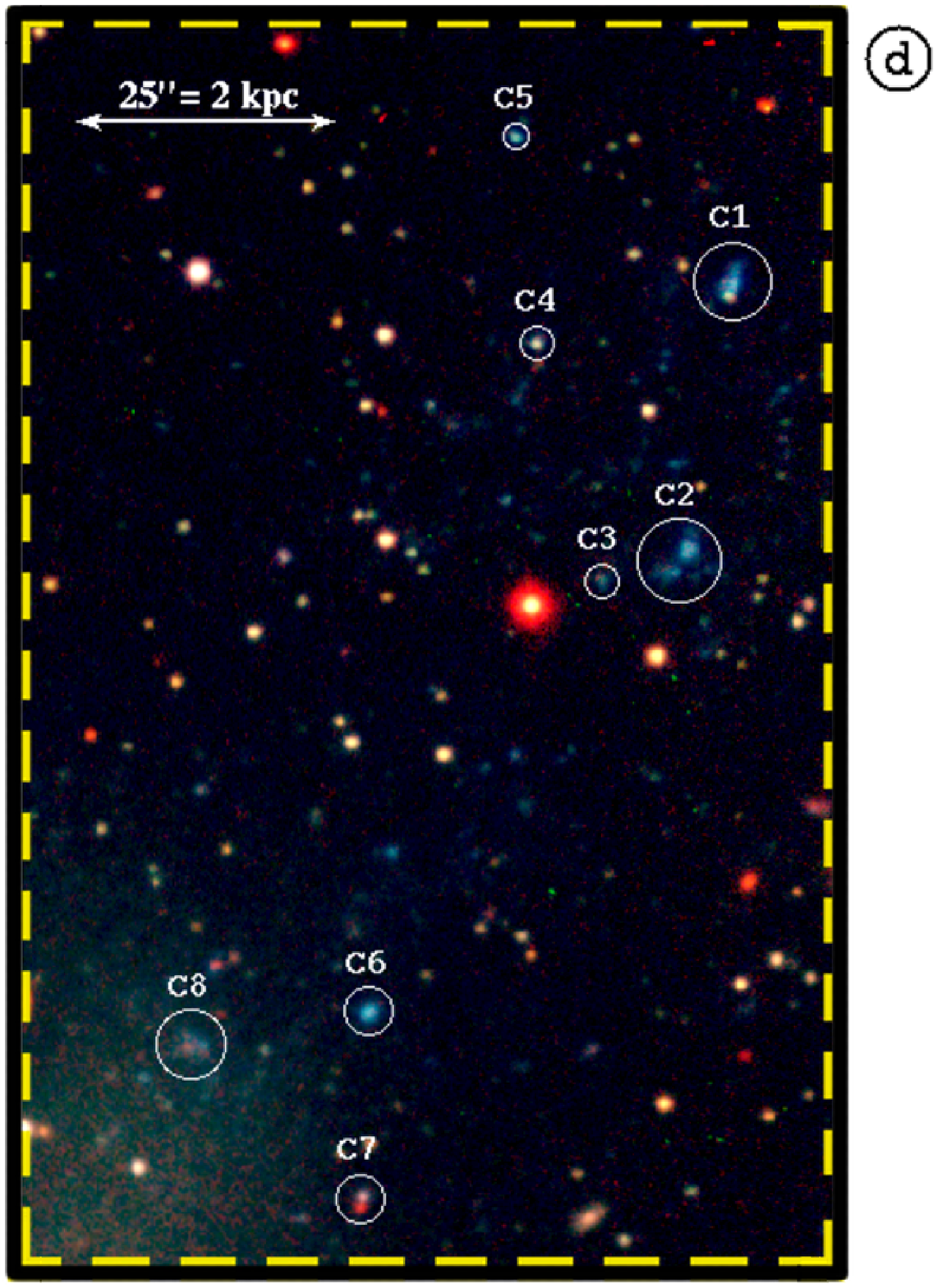} 
\end{tabular}											     	  
\caption{(a) RGB image of VCC1249 obtained combining 
the NGVS images in the $u, g, z$ filters. The outlying HII regions studied in this 
work are highlighted. Contamination from the light  associated with  the halo of M49 is 
clearly visible. (b) RGB image of VCC1249 obtained after subtracting a model for the 
light distribution in M49. 
(c) NUV contours superposed on the RGB image of VCC1249, after subtracting M49. 
(d) Enlargement of the region enclosed in the dashed yellow box in panel (b).
 Faint blue structures are visible near the highlighted HII regions. } 
	  \label{Fig2}														
\end{figure*}

\begin{table*}
\caption{Aperture photometry corrected for extinction in the Milky Way of the two galaxies and of the 
star-forming regions. The photometry of M49 is computed within an aperture of about 5.7 arcmin radius 
(or $\sim 1.6$ R$_e$; Ferrarese et al. 2006).}
\centering
{\footnotesize \begin{tabular}{ccccccccc}
\hline
\hline
 ID & Distance &   $u$    &    $g$  &   $i$   &   $z$   &   NUV  &   FUV  \\
  & (arcsec) &   (mag)    &    (mag)  &   (mag)   &   (mag)   &   (mag)  &   (mag)  \\
{\tiny(1)}  & {\tiny(2)} & {\tiny(3)} & {\tiny(4)}  & {\tiny(5)}  &  {\tiny(6)} &  {\tiny(7)} &  {\tiny(8)}  \\
 \hline
 \\
     VCC1249  &  0  &  15.60  $\pm$  0.03 & 14.73  $\pm$  0.03 &  14.19 $\pm$  0.03  & 14.11 $\pm$ 0.03 & 17.46 $\pm$ 0.15 & 18.68 $\pm$ 0.18\\
     M49 &  334.0  &  10.28  $\pm$  0.03 & 8.83  $\pm$  0.03 &  7.81 $\pm$  0.03  & 7.49 $\pm$ 0.03 & 13.91 $\pm$ 0.15 & 14.82 $\pm$ 0.17\\ 
\\        
     C1   &  151.4	&  21.33  $\pm$  0.04 &  21.31 $\pm$ 0.04  &  21.53  $\pm$ 0.04  &  22.12  $\pm$   0.04  &  20.77  $\pm$  0.16  &  20.80  $\pm$  0.29\\
     C2   &  126.1	&  21.31  $\pm$  0.04 &  21.41 $\pm$ 0.04  &  21.83  $\pm$ 0.04  &  22.51  $\pm$   0.05  &  21.52  $\pm$  0.16  &  21.09  $\pm$  0.32\\
     C3   &  119.4	&  23.53  $\pm$  0.06 &  23.27 $\pm$ 0.06  &  23.15  $\pm$ 0.05  &  23.00  $\pm$   0.06  &  -	   	&  -	     \\
     C4   &  137.9	&  23.09  $\pm$  0.05 &  22.56 $\pm$ 0.06  &  22.33  $\pm$ 0.04  &  22.28  $\pm$   0.04  &  (23.04  $\pm$  0.15)  &  (22.48  $\pm$  0.40)\\
     C5   &  156.4	&  22.91  $\pm$  0.05 &  22.69 $\pm$ 0.05  &  23.00  $\pm$ 0.05  &  23.28  $\pm$   0.05  &  -	   	&  -	    \\
     C6   &   70.7	&  21.92  $\pm$  0.04 &  21.94 $\pm$ 0.04  &  22.31  $\pm$ 0.04  &  22.10  $\pm$   0.04  &  21.37  $\pm$  0.16  &  21.17  $\pm$  0.33\\
     C7   &   55.1	&  23.17  $\pm$  0.06 &  22.87 $\pm$ 0.06  &  21.75  $\pm$ 0.04  &  21.22  $\pm$   0.03  &  -	     	&  -	    \\
     C8   &   61.1	&  22.31  $\pm$  0.06 &  22.15 $\pm$ 0.06  &  22.14  $\pm$ 0.05  &  21.69  $\pm$   0.04  &  21.43  $\pm$  0.16  &  20.98  $\pm$  0.31\\
     C12  &    5.8	&  22.17  $\pm$  0.06 &  21.48 $\pm$ 0.07  &  21.11  $\pm$ 0.06  &  21.10  $\pm$   0.05  &  -    &  -	    \\
     C17  &   19.2	&  23.94  $\pm$  0.09 &  22.81 $\pm$ 0.16  &  22.03  $\pm$ 0.06  &  22.08  $\pm$   0.06  &  -    &  -       \\
\hline
\end{tabular}
}
\\
\small{(1) Naming of the star-forming regions as in Lee et al. (1997), (2) projected distance from 
the center of VCC1249. The photometric uncertainties are the quadratic sum of the ZP error and the 
Poisson error.}
\label{Table2}
\end{table*}

The whole photometry was corrected for Galactic extinction assuming 
$E(B-V)=0.022$ from Schlegel, Finkbeiner \& Davis (1998) and the Milky 
Way extinction law such that $A_{FUV}=8.38E(B-V)$, $A_{NUV}=8.74E(B-V)$, 
$A_{u}=4.90E(B-V)$, $A_{g}=3.70E(B-V)$, $A_{i}=2.00E(B-V)$ and $A_{z}=1.45E(B-V)$.
Similarly, a 6 \% correction was applied to H$\alpha$ for extinction in the Milky Way.

\subsection{Optical imaging: NGVS}

The system VCC1249/M49 was observed in the optical bands ($u, g, i, z$) as part of the 
NGVS\footnote{https://www.astrosci.ca/NGVS/\\The\_Next\_Generation\_Virgo\_Cluster\_Survey/Home.html} 
under excellent seeing conditions ($\leq 1$ arcsec, see Table \ref{Table1}). 
The exposure times for each filter are listed in Table \ref{Table1}.
Images were reduced and calibrated using a dedicated data reduction pipeline (Ferrarese et al. 2011) 
designed explicitly to recover faint, diffuse surface brightness features, reaching a surface brightness 
limit of $\mu_g = 29 \, \rm mag~arcsec^{-2}$ (2$\sigma$ above the mean sky).
The mean global sky background around VCC1249 was estimated and subtracted using 
the GALPHOT tasks MARKSKY and SKYFIT. The former allows us to mark several rectangular 
regions on which the sky background is estimated; the latter computes the mean
sky background in each region using a sigma clipping algorithm, determines the average between 
the regions and subtracts this constant value from the image. The sky regions are 
selected to avoid the halo of M49 and other bright objects.

\begin{figure}
\begin{center}
\includegraphics[width=2.7cm]{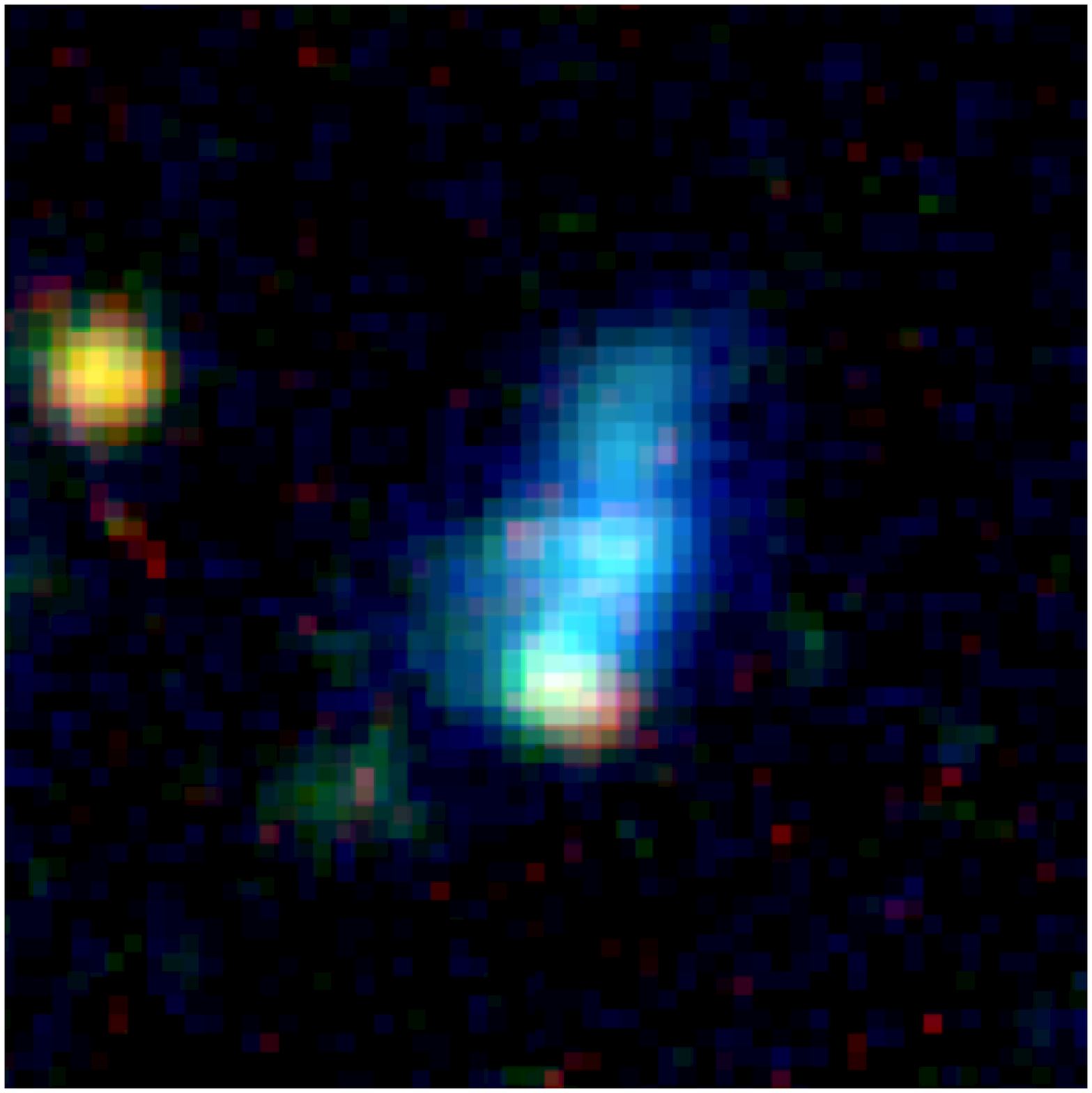} 
\includegraphics[width=2.7cm]{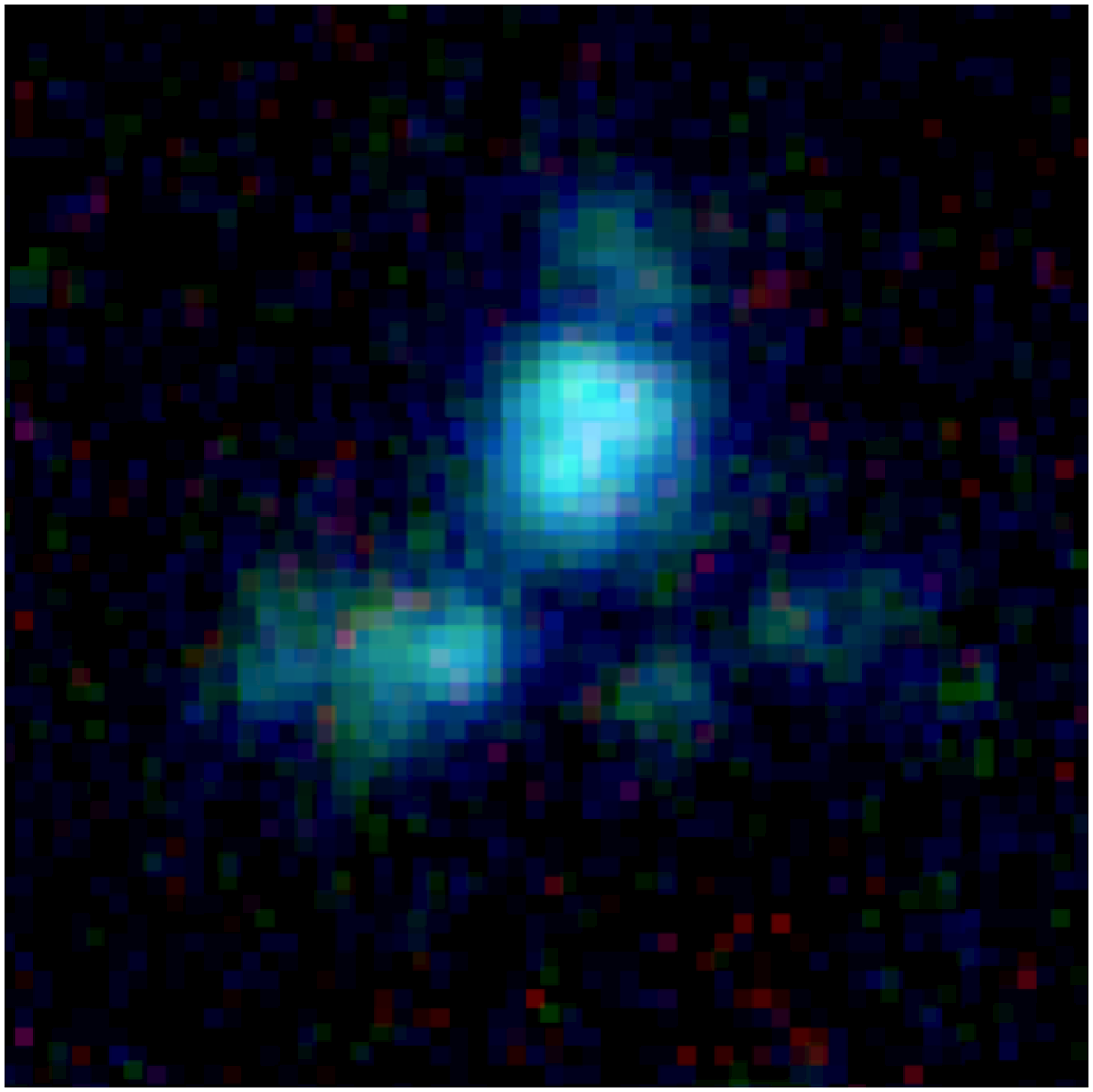} 
\includegraphics[width=2.7cm]{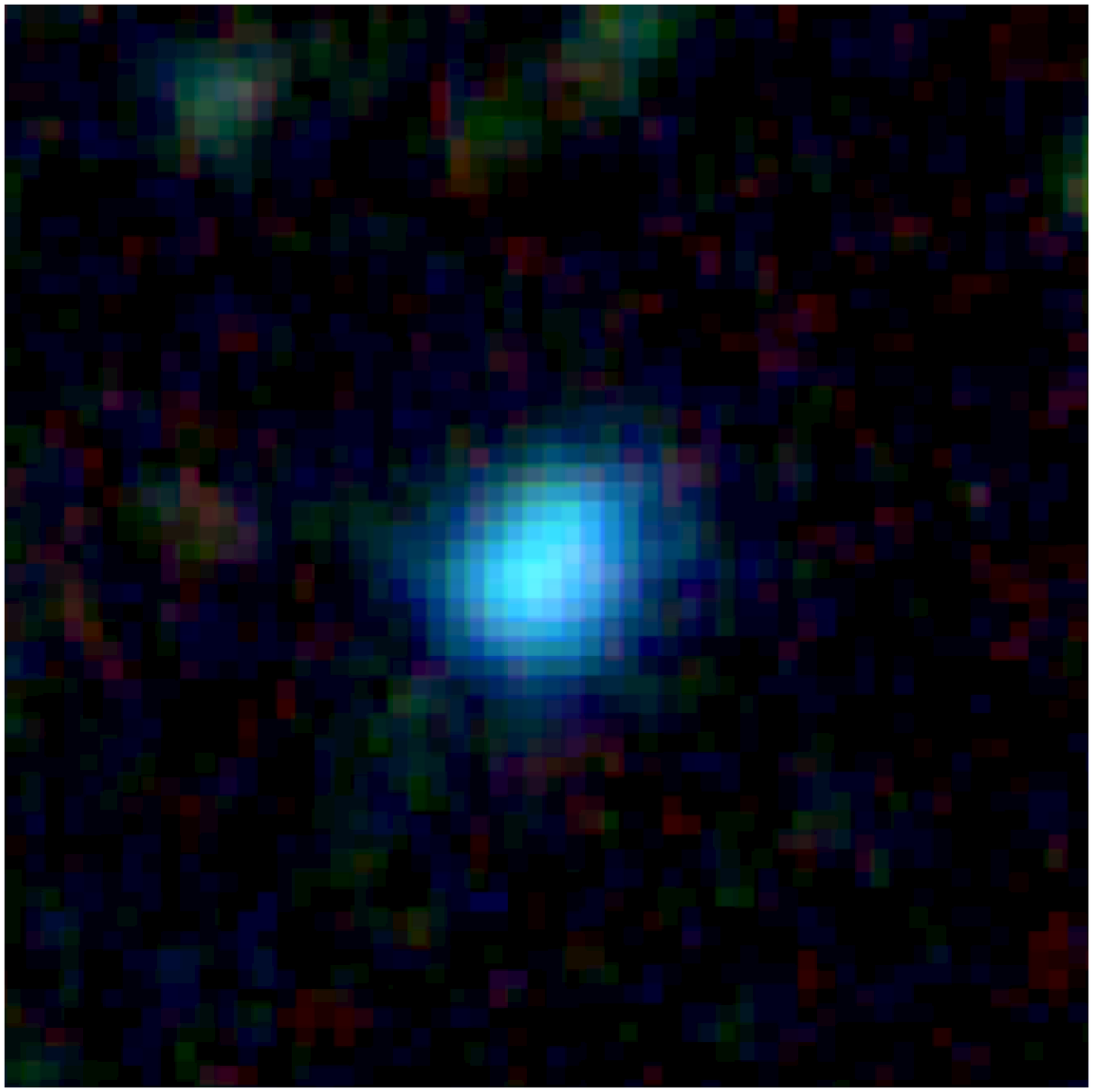}\\
\caption{RGB cutouts of the regions C1 (left), C2 (center), C6 (right) obtained 
combining the $u, g, z$ NGVS images. At the GALEX resolution of $>$ 4 arcsec,
C1 and C2 are unresolved in the UV. The size of each image is about $10''\times10''$.}
\label{Fig7}
\end{center}
\end{figure}

Because VCC1249 lies at $12^h30^m01^s.0 +07^d55^m46^s.0$ (J2000), only 5.6 projected arcmin from M49
(see Fig.\ref{Fig1}), it is difficult to derive accurate photometry of this galaxy
and its star-forming regions from the optical images, (in particular in the reddest filters) 
without first removing the contamination from the halo of the nearby giant elliptical. 
We therefore subtracted the best-fit model of the M49 light distribution from the sky-subtracted images, 
which we obtained for each filter ($u, g , i, z$) using in sequence the tasks ELLIPSE and BMODEL 
after masking VCC1249, the star-forming regions, and  the bright stars.
Panels (a) and (b) of Figure \ref{Fig2} show the RGB image before and 
after the subtraction of M49. To assess the quality of this subtraction, 
we analyzed in each subtracted image the background level in circular rings centered at the 
position of M49, finding no residual flux within the uncertainty of the sky level.

After removing the contamination from M49, we performed aperture 
photometry on the individual star-forming regions that are marked in Figure \ref{Fig2}.
The flux was measured in circular regions of about $3-4$ arcsec in radius, subtracting 
the local background determined within concentric annular regions of 5 to 10 arcsec in radius.
Throughout this work, we assume the nomenclature introduced by Lee et al (1997). 
Notice the presence of other fainter blue features near the listed 
HII regions (see panel (d) Figure \ref{Fig2}), likely low surface brightness star-forming 
filaments or knots related to the interaction studied. Owing to their faintness and uncertain flux, 
these will not be further analyzed.
For  VCC1249 we also measured the surface brightness profile in each band 
(after masking the foreground stars 
and the background objects), using a modified version of the ELLIPSE task in IRAF 
that fits elliptical isophotes. For consistency we performed the ellipse-fit on the 
$g$-band image and used the same elliptical isophotes for the other bands. 
During the analysis of the $g$-band image we maintained as free parameters 
the ellipse center, the ellipticity, and the position angle, and incremented the 
ellipse semi-major axis by 5 arcsec at each step of the fitting procedure. 

\subsection{UV imaging: GUViCS}

\begin{figure}[h!]
\begin{center}
\includegraphics[scale=0.59]{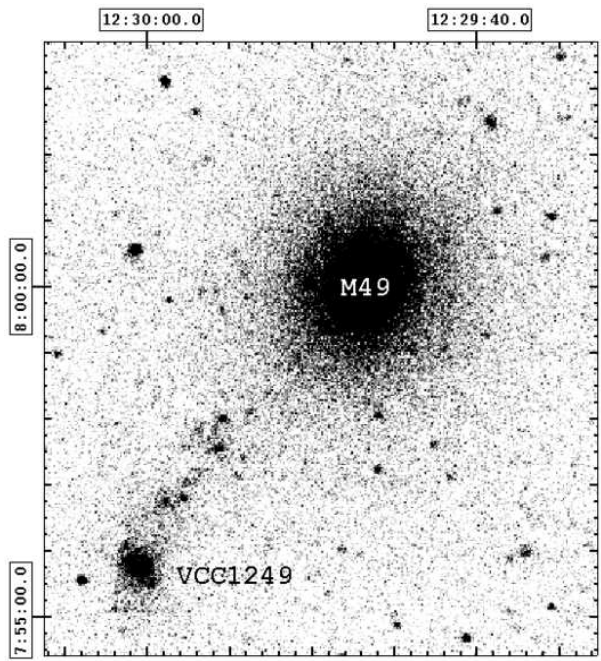} 
\caption{NUV image of VCC1249 and M49 obtained from GUViCS.  Multiple HII regions are trailing in the 
northeast direction from VCC1249.}
\label{FigNUV}
\end{center}
\end{figure}

The system of VCC1249 and M49 was observed by GALEX in the NUV (1750-2750 $\AA$) 
as part of the GUViCS\footnote{http://galex.oamp.fr/guvics/} survey
with an exposure time of 1630 sec. The NUV image, shown in Figure \ref{FigNUV}, 
reveals patchy extended emission between VCC1249 and M49. Compact emission is also seen 
in correspondence to some of the external HII regions. A shorter 104 sec exposure is 
available in the far-ultraviolet (FUV) band (1350-1750 $\AA$) from the GALEX all-sky survey. 
Owing to the shallower exposure, only VCC1249 and the brightest outlying HII regions 
(C1, C2, C6 and C8) are marginally detected in the smoothed data.
Because of the lower resolution of GALEX ($>$ 4 arcsec), regions C3, C5, and C7 are not detected in the UV images. 
Furthermore, some regions that are unresolved at the GALEX resolution (e.g. C2) appear to be blends of two or 
more distinct HII regions resolved in the optical imaging (Figure \ref{Fig7}). C1 is resolved into 
three components, one of which might be a background galaxy based on its red colors. 

Similarly to what was done for the optical imaging, we performed aperture photometry 
for the galaxy and the outlying HII regions. Owing to the lack of deep IR data, 
the internal dust extinction for VCC1249 is unknown. 
The IRAS 60 and 100 micron upper-limits (GoldMine\footnote{Gavazzi et al. (2003), 
Galaxy On Line Database Milano Network (http://goldmine.mib.infn.it)}) 
imply that $A(FUV)$ is lower than $\sim2$ mag (Cortese et al. 2008).
Moreover, the observed UV-optical colors of this galaxy are 
typical of blue-sequence dwarf 
galaxies (Boselli et al. 2008a; Cortese \& Hughes 2009), suggesting that the amount of dust absorption 
is likely modest ($<1$ mag). We tested the robustness of our
results against the adopted value of dust attenuation, finding that our
analysis is not affected if $A(FUV)<2$. Thus, for simplicity,
 we assume $A(FUV)$=0 throughout this paper.
For the HII regions instead we estimated the dust obscuration
using the Balmer decrement. Because the measured $I(H\alpha)/I(H\beta)$ 
ratio is 2.66 and 2.93 for C2 and C6, respectively, consistent with the 
expected Balmer ratio of 2.86
(case B for $T=10000$ K; Osterbrock 1989),  
no correction was applied.

\subsection{$H\alpha$ imaging}

\begin{figure}[h!]
\begin{center}
\includegraphics[scale=0.48]{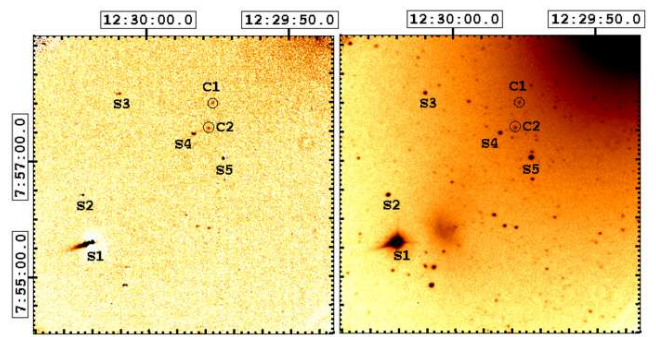} 
\caption{H$\alpha$ image of VCC1249 and the outlying HII regions obtained at the SPM telescope. 
Left: H$\alpha$ net emission line. Right: H$\alpha$ plus stellar continuum frame 
HII regions C1 and C2 are labeled. Stars with apparent residual NET emission (see text) are labeled S1 to S5.
Note the lack of substantial emission from VCC1249.}
\label{Ha}
\end{center}
\end{figure}

A 3$\times$600 sec narrow band $H\alpha$ image was acquired with the SPM\footnote{Observatorio Astron\'omico Nacional en 
San Pedro M\'artir, Baja California, Mexico.} 2.12m telescope in April 
2011 (see Table \ref{Table1}).These new data were combined with 
a 4$\times$300 sec exposure  acquired with the same instrument in 2001 that is available from the 
Goldmine database. A detailed description of the data reduction 
procedures for H$\alpha$ images can be found in Boselli \& Gavazzi (2002), and it is  only briefly summarized here. 
The ON-band frame was obtained using a narrow band (80 \AA) filter centered at 6603 $\AA$ that 
overlaps with the H$\alpha$ emission line. The OFF-band 
frame, needed to subtract the stellar continuum, was acquired in a shorter exposure using
the Gunn $r$-band filter. After bias subtraction and flat-fielding, the OFF band image was renormalized 
to match the ON-band exposure. Finally, the OFF-band was subtracted from the ON-band to obtain the 
image of the continuum-subtracted NET H$\alpha$ emission.
The resulting NET and ON-band frames are shown in Figure \ref{Ha}. 
Although the normalization factor was determined assuming that 
no H$\alpha$  flux should be found associated to stars in 
the field, some residual flux from the brightest stars in the field 
appears in the NET frame due to slight differences in
seeing between the OFF and ON observations (see Figure \ref{Ha}). 
The absolute flux calibration was performed with repeated exposures of the standard 
spectrophotometric stars Feige34 and Hz44.

In the final image, two sources (C1 and C2) with an H$\alpha$ surface 
brightness higher than $\sigma_{min} = 2.0 \times 10^{-17} $ erg cm$^{-2}$ s$^{-1}$ arcsec$^{-2}$ 
were detected (5$\sigma$ above the background). As previously, we measured the flux in circular apertures 
obtaining  $(8.68\pm0.33) \times 10^{-16}$ erg 
cm$^{-2}$ s$^{-1}$ and $(1.49\pm0.04) \times 10^{-15}$ erg cm$^{-2}$ s$^{-1}$ for C1 
and C2, respectively. 
Since the narrow band filter is broad enough 
to include the bracketing
[NII] lines, the listed H$\alpha$ fluxes were corrected accordingly, 
using $[NII]/H\alpha=0.11$ for C2, as measured on the spectrum, while for C1 we used 
an average ratio of about 0.1, as derived from the spectra of C2 and C6.

\begin{figure*}
\centering
\includegraphics[scale=0.8]{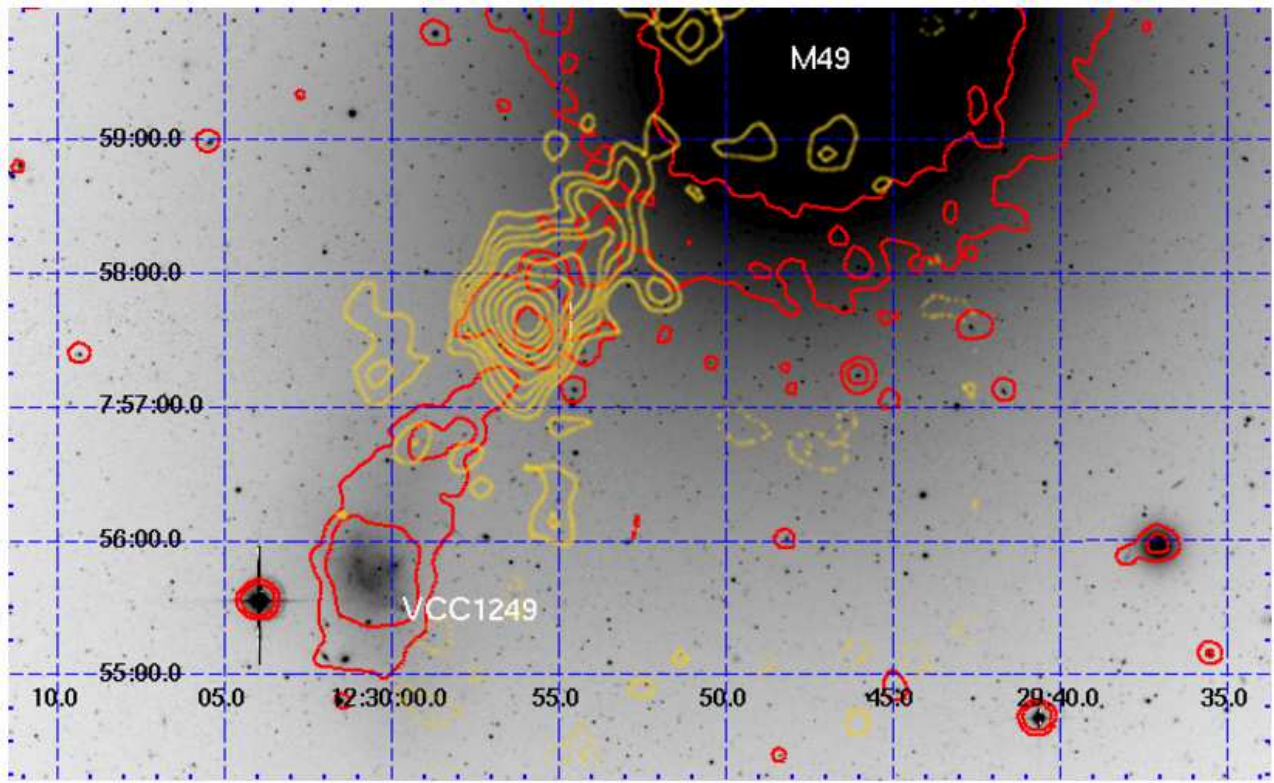} 
\caption{NGVS g image of VCC1249 (bottom-left) and of M49 (top-center) on which the NUV contours 
(red) and the HI contours taken from McNamara et al. (1994) (yellow) are superposed (the coordinates 
are precessed from B1950 originally used, to J2000). Note that the peak of the HI cloud nearly coincides 
with the peak of the NUV emission at the position of the region C2 (LR1 in Lee et al. 2000).}
\label{withHI}
\end{figure*}

\subsection{Spectroscopy}

We observed the HII regions C2\footnote{LR1 in Lee et al. 2000}, C4, and C6 on UT 2011 January, 3 and 4
with the LRIS spectrograph (Oke et al, 1995) at the Keck I telescope on Mauna Kea. 
During the first night we obtained $2\times 1200$ sec exposures of C2 and C4, 
using a $1''$ slit and with the instrument configured with the 600/4000 grism, 
the D560 dichroic and the 600/7500 grating, tilted to ensure continuous spectral 
coverage between the blue and red arms.
A single exposure of 900 sec was obtained during the second night, 
with a similar configuration but using the 400/3400 grism.
The 2D images were reduced with the LowRedux 
pipeline\footnote{http://www.ucolick.org/∼xavier/LowRedux/index.html}, 
which calibrates, extracts and fluxes the data.
C1 was subsequently observed on the UT night 2011 April, 29 with the KAST spectrograph at 
the Lick observatory. We obtained $2\times 1800$ sec exposures with the
600/4310 grism, the 600/7500 grating and the D55 dichroic using a $2''$ slit.
For regions C12 and C17 the spectra taken in $3''$ apertures
were retrieved from SDSS. 

\section{Properties of VCC1249}
\label{Galaxy Morphology}

VCC1249 is a dIrr (Nilson et al. 1973) low-surface brightness galaxy, located 5.6 arcmin to the southeast of the 
giant elliptical galaxy M49 (Kumar \& Thonnard 1983), the brightest member of  Virgo Cluster B, approximately 
4 degrees (1.19 Mpc projected distance) south of M87 (Virgo Cluster A). 
VCC1249 has an irregular morphology: the central part of the galaxy consists of several 
bright knots connected by bridges (Lee et al. 1997) and from this structure
a tidal tail departs toward  M49 in the northwest direction and a counter tail in the southwest 
direction (Patterson \& Thuan 1992 and McNamara et al. 1994; see Figure \ref{Fig2}).
Based on its photometric properties (Table \ref{Table2}), VCC1249 is
a dwarf galaxy that lies in the blue cloud of the color-magnitude diagram of the Virgo cluster
galaxies. 

VCC1249 has a systemic velocity of 390$\pm$30 km s$^{-1}$ (SDSS)
\footnote{This is the redshift reported by SDSS for regions C12 and C17. 
Huchra (1992) found 276$\pm$78 km s$^{-1}$. Using this value, the relative velocity 
between VCC1249 and M49 would increase and the ram pressure would appear more effective. 
In turn, this would lead to a more rapid removal 
of the gas (58$\pm$21 Myr), leaving our conclusions unchanged, however.}, 
while M49  has
1001$\pm$12 km s$^{-1}$ (Schechter 1980). 
An HI cloud with $M_{HI}=(6.9\pm0.4) \times 10^7$ M$_{\odot}$ and radial velocity 
of 469 km s$^{-1}$ (McNamara et al. 1994), in agreement with the previous measurement 
of 472$\pm$4 km s$^{-1}$ (Sancisi et al. 1987), 
belonging to VCC1249 and displaced toward M49, is detected in 21-cm (see Figure \ref{withHI}).
Conversely,  VCC1249 itself shows no significant emission in the hydrogen line to a limit of 
$M_{HI}<4.2 \times 10^7$ M$_\odot$ (Oosterloo \& Shostak 1984).
Quantified in terms of the HI deficiency ($def_{\rm HI}$), defined as
the logarithmic difference between the upper limit HI mass and the expected value
for a galaxy of the same morphological type and size (Haynes \& Giovanelli 1984), 
VCC1249 has $def_{\rm HI}>0.92$ and is among the most deficient galaxies
in the local Universe. This is not particularly surprising, since at the projected distance 
between M49 and VCC1249 of about 30 kpc, the X-ray emission is dominated by 
the hot ($10^7$ K  Forman et al. 1985) gas belonging 
to the halo of M49 with a density of more than $10^{-3}$ cm$^{-3}$ (Fabian 1985).

\begin{figure}
\begin{center}
\includegraphics[width=9.0cm]{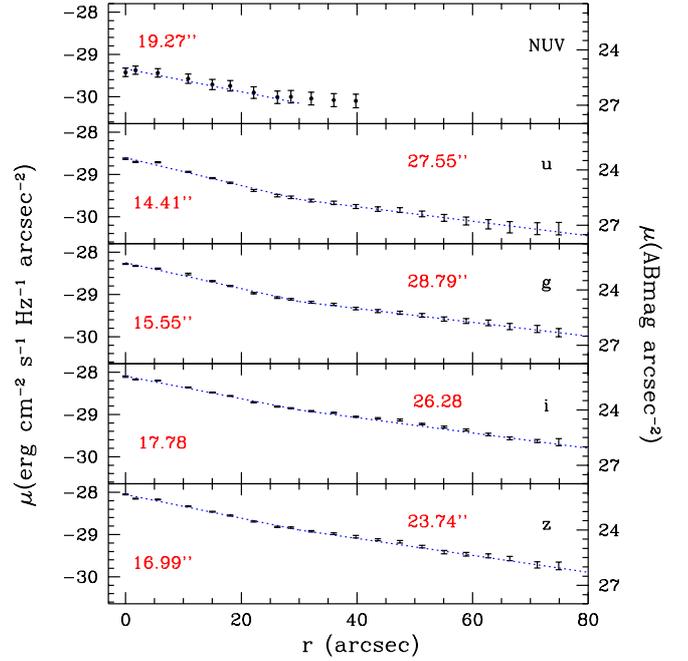}\\ 
\caption{Surface brightness profile of VCC1249, obtained using ELLIPSE, letting the center, 
the ellipticity, and the position angle as free parameters.
In red, we show the inner and outer scale lengths computed for each band adopting a two-component exponential profile
(superimposed in blue). $r = \sqrt{a \cdot b}$, where $a$ and $b$ are 
the semi-major and the semi-minor axis of the isophotal ellipses, respectively. 
}
\label{Fig5}
\end{center}
\end{figure}

In Figure \ref{Fig5} we show the surface brightness profile of VCC1249 
as a function of the radius $r = \sqrt{a \cdot b}$, where $a$ and $b$ are 
the semi-major and the semi-minor axis of the isophotal ellipses, respectively.
The profiles are well fitted by an exponential law with two 
components (for $0'' < r < 30''$ and $ 30''< r < 80'' $). 
The first component is characteristic of a disk,  while the second one accounts for 
the presence of the tidal tail and its counter-tail (Patterson \& Thuan 1992; Lee et al. 1997),
which appear as an excess of light above the inner exponential fit, 
resulting in a greater scale length in the outer part of the profile.

\begin{figure}[!t]
\begin{center}
\includegraphics[width=8.8cm]{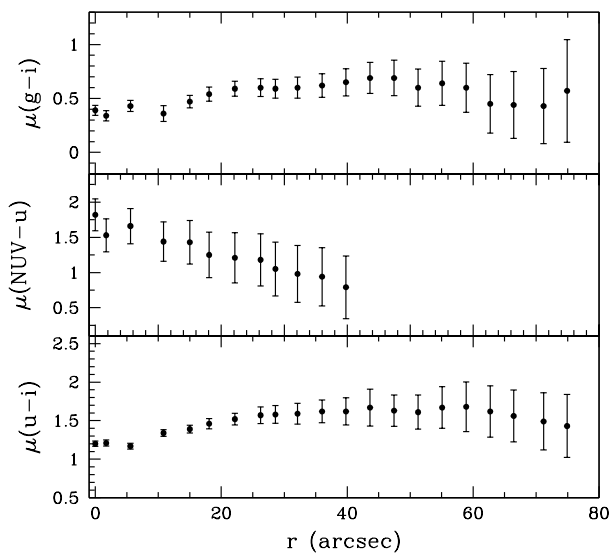}\\ 
\caption{Color profile of VCC1249, obtained from the surface brightness profiles in Figure \ref{Fig5}.}
\label{Fig6}
\end{center}
\end{figure}

The color profiles of VCC1249 are shown in Figure \ref{Fig6}. The $u-i$ and $g-i$ indices are as blue as
$u-i=1.20$ and $g-i=0.39$ near the center, become redder moving outward up to 20 arcsec, then they flatten, consistently
with the results of Lee et al. (1997) and Patterson \& Thuan (1996). 
However, the $NUV-u$ color shows an opposite trend: from $NUV-u$=1.8 near the center to 0.8 at 40 arcsec. Note, however, 
that the uncertainties in these colors are large.

Inspecting our H$\alpha$ imaging,  we see a lack of significant H$\alpha$ emission
associated to VCC1249 and to its tidal tails. Using an upper limit of 
F$_{H\alpha}^{VCC1249}<3.80\times10^{-14}~\rm erg\phantom{x}cm^{-2}\phantom{x}s^{-1} $, 
we infer an upper limit to the star formation rate (SFR) of about $<0.01~\rm M_{\odot}\phantom{x}yr^{-1} $, 
using the calibration of Kennicutt (1998). Because H$\alpha$ traces massive star formation, the lack of 
significant H$\alpha$ emission sets an upper limit on the age of the last star-formation event 
at $\gtrsim 20$ Myr, i.e. the typical 
life-time of massive stars. Using the FUV emission instead, we infer an SFR of about 
$0.005\pm0.001 ~\rm M_{\odot}\phantom{x}yr^{-1} $ following Kennicutt (1998). 
The presence of far-ultraviolet continuum indicates that star formation ocurred $\gtrsim 100$ Myr 
ago, which combined with the lack of H$\alpha$ emission, suggests that the star formation in 
VCC1249 was truncated recently. 

\section{Properties of the external star-forming regions}
\subsection{Morphology and photometry}
\label{MandPregions}

A complex of star-forming regions
extends from the galaxy in northwest direction, up to 2.6 arcmin 
(about 13 kpc) far from VCC1249, as clearly visible from both the GALEX images
and the NGVS images (Figure \ref{Fig2}a,b). 

\begin{figure}[h!]
\centering
\includegraphics[width=9cm]{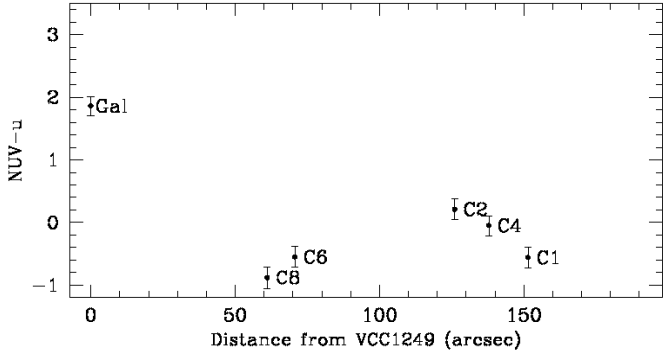} 
\caption{$NUV-u$ color of the studied HII regions and of VCC1249 itself as a function of the 
projected distance from VCC1249. All outlying HII regions are bluer than VCC1249.
} 
\label{Fig8}
\end{figure}

Figure \ref{Fig8} shows the $NUV-u$ colors of these star-forming regions compared to VCC1249. 
It appears that all the structures outside VCC1249 are much bluer than the galaxy 
($NUV-u=1.86$ and $u-i=1.41$). C12 and C17, 
belonging to the central region of VCC1249, are unresolved in the UV images and appear redder 
in the $u-i$ color than all other regions, having 
$\left(u-i\right)_{C12}= 1.06$ and $\left(u-i\right)_{C17}= 1.91$, respectively. 
Conversely, the other regions have a mean color of  $\langle NUV-u \rangle= -0.29$ and 
$\langle u-i \rangle =0.21$. Only  region C7 is as red as the regions belonging to the galaxy in $u-i$, 
but visual inspection of the imaging suggests that this is probably due to contamination from a 
background red source (see Figure \ref{Fig2}). Furthermore, the lack of a clear dependence of these colors as a function 
of the projected distance from the galaxy suggests that these regions
are nearly coeval, regardless of the galactocentric distance.  

Only regions  C1 and C2 display H$\alpha$ emission at the depth of our imaging. 
Their luminosities are $2.78  \times 10^{37} \rm erg\phantom{x}s^{-1} \rm$  and 
$4.78  \times 10^{37} \rm erg\phantom{x}s^{-1} \rm$, respectively, consistent
with the faint end of the HII luminosity function (Kennicutt et al. 1989). 
Adopting the Kennicutt (1998) conversion, the inferred SFR is 
$2.20 \times 10^{-4} ~\rm M_{\odot}\phantom{x}yr^{-1} \rm$ (C1) 
and $3.77 \times 10^{-4} ~\rm M_{\odot}\phantom{x}yr^{-1} \rm$ (C2).
Using the FUV emission instead, we compute an SFR of about $(7\pm2) \times 10^{-4}~\rm M_{\odot}\phantom{x}yr^{-1} $ 
and $(6\pm2) \times 10^{-4}~\rm M_{\odot}\phantom{x}yr^{-1} $, for C1 and C2, respectively. 
The SFR of the other regions can be estimated from the FUV luminosity to approximately 
$10^{-4}~\rm M_{\odot}\phantom{x}yr^{-1} \rm$ per region (see Table \ref{Table8})\footnote{
The Kennicutt (1998) calibration applies to continuous star formation over $10^8$ yr or longer. 
Therefore, the estimated SFR of the HII regions should be approached with some caution.
}. 

\subsection{Radial velocities and line properties}

\begin{figure*}[ht!]
\centering
\vspace{-3.7cm}
\includegraphics[scale=0.715]{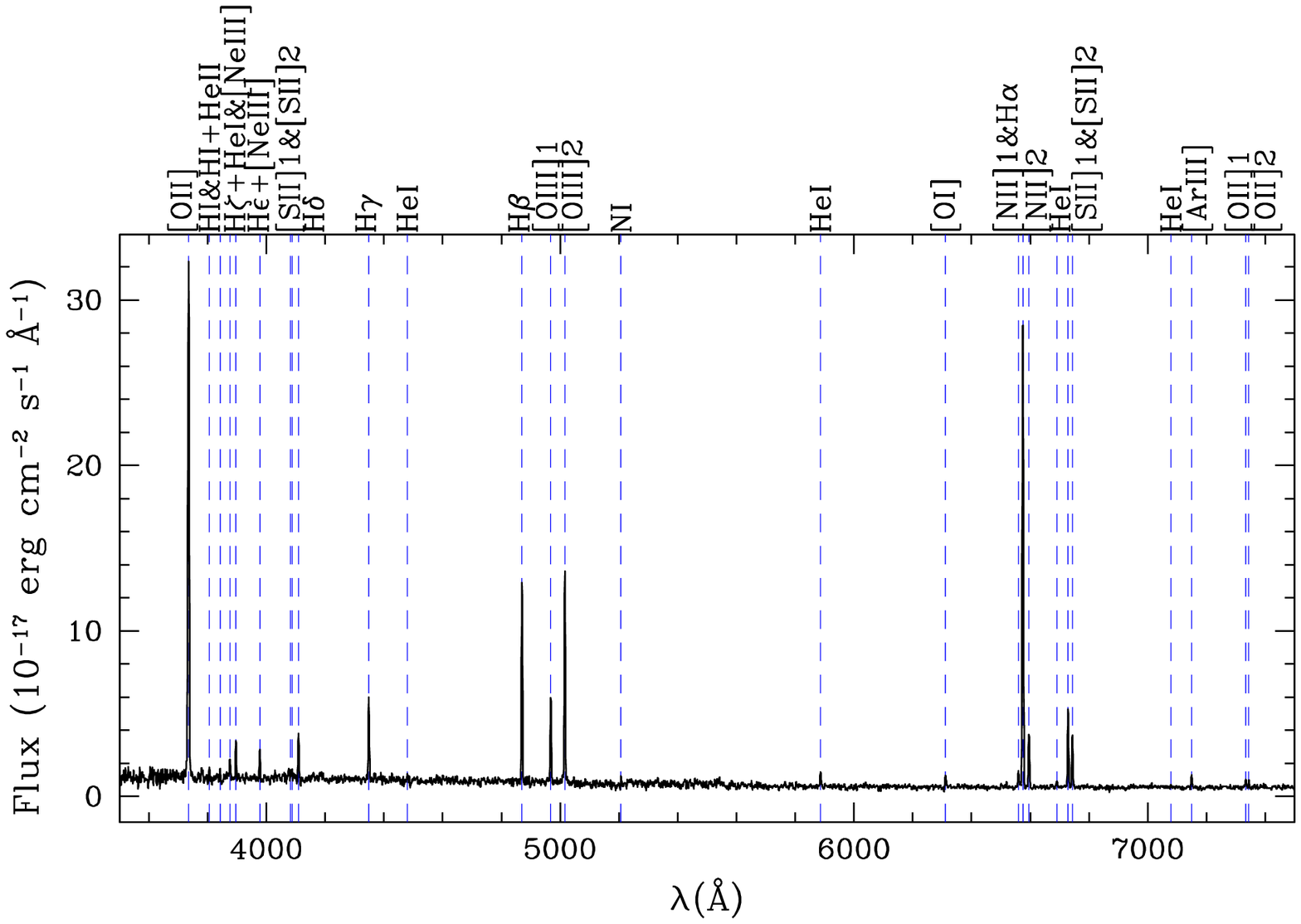}\\ 
\vspace{-6cm}
\includegraphics[scale=0.715]{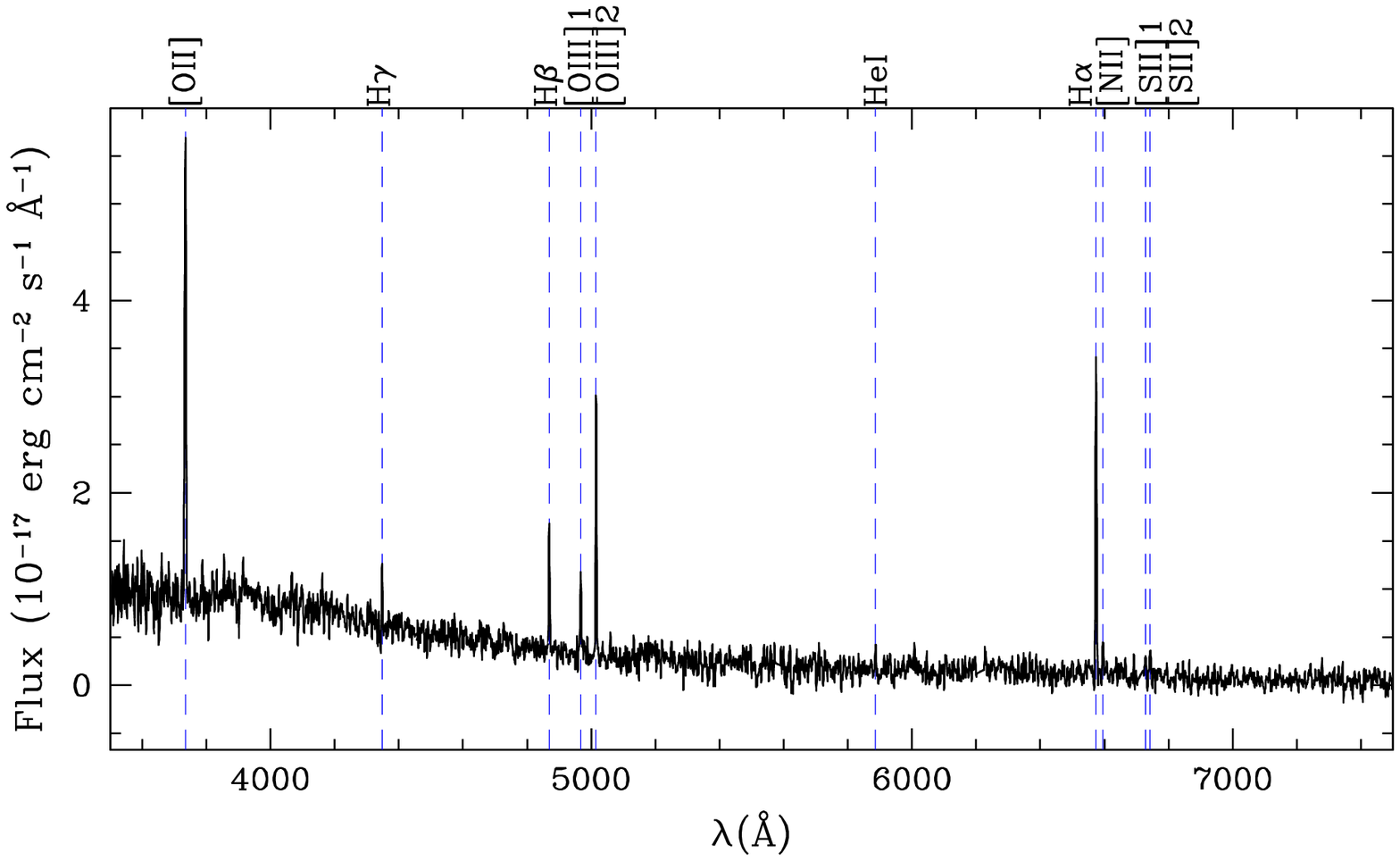}
\caption{Spectrum of the HII regions C2 (top) and C6 (bottom) obtained at Keck. 
Dashed lines highlight the position of the lines in Table \ref{Table6}.}
\label{Fig9}
\end{figure*}

The Keck spectra of C2 and C6 are shown in Figure \ref{Fig9}, where we highlight the detected 
emission lines. These two spectra have a mean signal-to-noise ratio (S/N) of 3.7 and 2.2 \footnote{Obviously, the S/N turns out to be much higher 
in correspondence of the lines, exceeding 10 for the C2 and C6 regions.}, respectively. 
The lower S/N in the spectra for the fainter HII regions C4 and C1, 0.9 and 1.4, respectively, 
prevented a detailed study of their emission properties, but we can derive an estimate of the 
radial velocity from the marginally detected H$\alpha$ lines. Since the measured $I(H\alpha)/I(H\beta)$ 
ratio is about 2.66 and 2.93 for C2 and C6, respectively, consistent with the expected Balmer ratio of 2.86
(case B for $T=10000$ K; Osterbrock 1989) within the uncertainty,  
no reddening correction was applied to the spectra.

We computed the radial velocities of C2 and C6 by averaging the redshift measurements 
obtained from the individual spectral lines. C2 and C6 have a radial velocity 
of 561$\pm$34 km s$^{-1}$ and 533$\pm$53 km s$^{-1}$, respectively.
Our redshift measurement for C2 agrees with 
the value $577\pm91 ~\rm km\phantom{x} s^{-1}$ from Lee et al. (2000). 
The radial velocities listed in Table \ref{Table4} systematically increase 
toward M49 (1001$\pm$12 km s$^{-1}$). HI interferometric observations carried out with the VLA have shown 
that regions C1-C5 are apparently associated with an HI cloud of $\simeq$ 6.9$\times$10$^{7}$ M$\odot$ at a 
recessional velocity of 469 km s$^{-1}$ extending over 20 km s$^{-1}$ (McNamara et al. 1994). The VLA channel maps 
show a velocity gradient across the HI cloud: the material in the northwest part has higher velocities (10-15 km s$^{-1}$) 
compared to the center (McNamara et al. 1994).  
Having a bandpass of about 330 km s$^{-1}$ centered at 470 km s$^{-1}$, these observations (as well as the 
WSRT observation by Henning et al. 1993) cannot be 
used to check if the association on the plane of the sky is also present in the velocity space.
Lower resolution WSRT observations (Sancisi et al. 1987) and Arecibo observations (Hoffman et al. 1987) covering a 
broader range in the velocity space, however, do not detect any 21 cm emission 
with an rms of 0.7 mJy at the redshift of C1 or C4, and probably C2 and C6 (see Table 3).

\begin{table}[h!]
\caption{Radial velocities for C1, C2, C4, C6, C12, and C17.}
\centering
{\footnotesize \begin{tabular}{c|c|c}
\hline
\hline
  Region 	& Proj.Dist.   & Velocity       \\
   	   & (kpc) & (km s$^{-1})$	      \\
\hline
(VCC1249)  &	0     & 390$\pm$30      \\
C12 	   &	0.5   & 390$\pm$30      \\
C17 	   &   1.5    & 390$\pm$30      \\
C6  	   &   5.6    & 533$\pm$53      \\
(HI)	   &  10.0    & 469$\pm$3      \\
C2   	   &  10.0    & 561$\pm$34      \\
C4  	   &  11.0    & 656$\pm$73     \\
C1         &  12.0    & 716$\pm$106     \\
(M49)	   &  26.5    & 1001$\pm$12     \\
\hline
\end{tabular}
}
\label{Table4}
\end{table}

The observed equivalent widths (EW) of the lines measured in C2 and C6 
(adopting the convention that negative EW means emission) are listed in Table \ref{Table6}. 
From the ratios $\left[OIII\right]\lambda5007$/H$\beta$ 
and $\left[NII\right]\lambda6584$/H$\alpha$, we determine that, as expected,  C2 and C6 
lie on the HII regions arm of the BPT diagram (Baldwin et al. 1981). No clear features related to 
shocks are seen in the spectra: no strong $[OI]$  line is detected and all diagnostic 
ratios (involving $[OI]$, $\left[OIII\right]$, H$\beta$, H$\alpha$, $\left[SII\right]$, $\left[NII\right]$ lines) 
agree with the photo-ionization models that place C2 and C6 among HII regions 
(see e.g. Dopita \& Evans 1986). 

Conversely, the SDSS spectra of C12 and C17 belonging to VCC1249 
exhibit no emission lines, but clearly show Balmer lines in absorption. In Table \ref{Table5} 
we report the line equivalent widths. Consistent with our analysis of the star formation rate properties, 
the observed values are typical of k+a galaxies (EW(H$\delta$)$> $3 \AA, Poggianti et al. 2004, 
Dressler et al. 1999), post star-burst (PSB) galaxies whose star formation
has been suddenly truncated. These characteristics are not only found in C12 and C17, but are 
representative of the entire galaxy. Indeed, the spectrum of VCC1249 (Patterson \& Thuan 1992;
Huchra 1992) exhibits no emission lines and strong H$\beta$ in absorption.

\begin{table}[h!]
\caption{The Balmer absorption lines of C12 and C17.}
\centering
{\footnotesize \begin{tabular}{c|c|c|c}
\hline
\hline
   	    & Wavelength & EW  (C12) & EW (C17)   \\
   	    & ($\AA$)	   & ($\AA$)& ($\AA$) \\

\hline
H$\delta$   & 4101.7	   &   5    &  4    \\
H$\gamma$   & 4339.8	   &   6    &  2   \\
H$\beta$    & 4861.0	   &   5    &  4    \\
H$\alpha$   & 6562.8	   &   1    &  7   \\
\hline
\end{tabular}
}
\label{Table5}
\end{table}

\begin{table*}
\caption{Summary of the emission lines detected in the HII regions C2 and C6}
\centering
{\footnotesize \begin{tabular}{c|c|ccc|ccc}
\hline
\hline
	    &				   &   & C2 & &    &C6&  \\                           
   	    & $\lambda$ (rest) & $\lambda$ (observed) & Log Continuum & EW & $\lambda$ (observed) & Log Continuum &      EW     \\
   	    & ($\AA$)	      & ($\AA$)		&  (erg cm$^{-2}$ s$^{-1}$ $\AA^{-1}$)   & ($\AA$)    & ($\AA$) & (erg cm$^{-2}$ s$^{-1}$ $\AA^{-1}$) & ($\AA$)  \\
\hline
 & & & & & & & \\
$\left[OII\right]$	    	& 3727.55     &   3735.2       &       -16.93	       & -148.9 $\pm$ 1.1    & 3734.8    & -17.04     & -29.5 $\pm$ 0.8	 \\
HI	    			& 3798        &   3805.9       &       -16.95	       & -1.7 $\pm$ 0.4      &  -	 &    -       & -		   \\
HI$+$HeII   			& 3834	      &   3842.4       &       -16.95	       & -1.4 $\pm$ 0.5      &  -	 &    -       & -		   \\
$\left[NeIII\right]$     	& 3868.7      &   3876.2       &       -16.95	       & -3.9 $\pm$ 0.4      &  -	 &    -       & -		   \\
H$\zeta+$HeI  			& 3888.9      &   3896.4       &       -16.96	       & -7.3 $\pm$ 0.4      &  -	 &    -       & -		   \\
H$\epsilon+\left[NeIII\right]$  & 3970.1      &   3977.6       &       -16.97	       & -6.3 $\pm$ 0.3      &  -	 &    -       & -		   \\
$\left[SII\right]$1      	& 4070        &   4082.9       &       -16.93	       & -0.7 $\pm$ 0.3      &   -	 &    -       & -		   \\
$\left[SII\right]$2	    	& 4078        &   4087.1       &       -16.93	       & -0.9 $\pm$ 0.3      &   -	 &    -       & -		   \\
H$\delta$   			& 4101.7      &   4109.6       &       -16.95	       & -9.8 $\pm$ 0.5      &   -	 &    -       & -		   \\
H$\gamma$   			& 4339.8      &   4348.7       &       -16.98	       & -19.8 $\pm$ 0.6     &  4348.3   & -17.21     & -3.4 $\pm$ 0.5	 \\
HeI	    			& 4471.5      &   4480.0       &       -17.02	       & -1.5 $\pm$ 0.4      &   -       &    -	    & -              	 \\
H$\beta$    			& 4861.0      &   4870.2       &       -17.06	       & -57.2 $\pm$ 0.6     &   4869.4  & -17.43     &  -13.6 $\pm$ 0.9     \\
$\left[OIII\right]$1     	& 4958.9      &   4967.9       &       -17.06	       & -24.5 $\pm$ 0.6     &   4967.3  & -17.49     &  -10.6 $\pm$ 1.1     \\ 
$\left[OIII\right]$2     	& 5006.8      &   5015.8       &       -17.07	       & -62.4 $\pm$ 0.8     &   5015.2  & -17.52     &  -34.3 $\pm$ 1.2     \\ 
NI	    			& 5199        &   5207.4       &       -17.17	       & -2.4 $\pm$ 0.6      &   -	 &    -       & -		   \\
HeI	    			& 5875.6      &   5886.3       &       -17.23	       & -6.6 $\pm$ 0.7      &   5886.5  & -17.82     & -5.9 $\pm$ 2.7	   \\
$\left[OI\right]$	    	& 6300.3      &   6311.7       &       -17.25	       & -5.9 $\pm$ 0.6      &   -	 &    -       & -		   \\
$\left[NII\right]$1	    	& 6548.1      &   6559.8       &       -17.28	       & -9.7 $\pm$ 0.6      &   -	 &    -       &  -		   \\
H$\alpha$   			& 6562.3      &   6574.8       &       -17.28	       & -252.9 $\pm$ 1.0    &   6574.1  & -17.89     & -115.2 $\pm$ 1.5	 \\
$\left[NII\right]$2	    	& 6583.4      &   6595.5       &       -17.28	       & -28.8 $\pm$ 0.5     &   6595.3  & -17.89     & -10.8 $\pm$ 2.1	 \\
HeI	    			& 6678        &   6690.5       &       -17.25	       & -3.2 $\pm$ 0.5      &   -       &    -	    &  -             	 \\
$\left[SII\right]$	    	& 6717        &   6728.8       &       -17.25	       & -38.8 $\pm$ 0.6     &   6727.9  & -18.00     &  -11.4 $\pm$ 2.7	 \\
$\left[SII\right]$	    	& 6731.3      &   6743.1       &       -17.25	       & -25.7 $\pm$ 0.6     &   6742.6  & -18.05     &  -17.1 $\pm$ 1.9	 \\
HeI				& 7065	      &   7078.4       &       -17.26	       & -0.9 $\pm$ 0.4      &   -	 &    -       & -		   \\
$\left[ArIII\right]$	    	& 7135.8      &   7149.1       &       -17.28	       & -6.3 $\pm$ 0.5      &   -	 &    -       & -		   \\
$\left[OII\right]$1	    	& 7319.5      &   7333.5       &       -17.27	       & -4.0 $\pm$ 0.4      &   -	 &    -       & -		   \\
$\left[OII\right]$2		& 7330.5      &   7344.1       &       -17.27	       & -3.0 $\pm$ 0.6      &   -	 &    -       &  -		   \\
\hline
\end{tabular}
}
\label{Table6}
\end{table*}

\subsubsection{Metallicity}

Since the $\left[OIII\right] \lambda4363$ line is undetected in the spectra of C2 and C6, we determined 
the oxygen abundances using the strong line methods.  Several such methods, 
based on different line ratios with empirical, theoretical or "combined" calibrations, 
can be found in the literature. 

\begin{table}[h!]
\caption{Oxygen abundances of the HII regions C2 and C6. The values converted to the Pettini \& Pagel (2004) 
calibration are given in parenthesis.}
\centering
{\footnotesize \begin{tabular}{c|c|c|l}
\hline
\hline
	& C2 & C6 & \\
Ratio & $12+log\left(\frac{O}{H}\right)$ & $12+log\left(\frac{O}{H}\right)$ & Calibration\\
\hline
                                     	    &	 		        &		             &		                   \\
$R_{23},O_{32}$                      	    &	 	8.69   (8.49)	& 8.41    (8.34)             & {\tiny McGaugh (1991)}		   \\
                                     	    &	 	                &	      		     &			           \\
$\frac{\left[NII\right]}{\left[OIII\right]}$&	 	8.45   (8.24)	& -	  		     & {\tiny Kewley \& Dopita (2002)}     \\
                                     	    &	 	        	&	   		     &  	                   \\
$R_{23},O_{32}$                             &	 	8.85   (8.48)	& 8.50    (8.19)	     & {\tiny Kobulnicky \& Kewley (2004)} \\
                                     	    &	 	        	&	   		     &  	                   \\
$R_{23}$                                    &	        8.81   (8.49)	& 8.74    (8.42)	     & {\tiny Zaritsky et al. (1994)}      \\
                                            &  		        	&	   		     &  	                   \\
$\frac{\left[NII\right]}{H_{\alpha}}$       &		8.31    	& 8.25     		     & {\tiny Pettini \& Pagel (2004)}     \\
                                            &  		        	&	   		     &  	                   \\
$\frac{\left[NII\right]}{H_{\alpha}}$       &		8.43   (8.31)	& 8.35    (8.25)	     & {\tiny Denicol\'o et al. (2002)}    \\
                                            &  			        &			     &		                   \\
\hline
\end{tabular}
}
\label{Table7}
\end{table}

Since the abundances derived using these indirect methods 
are affected by up to $\sim0.7$ dex uncertainties (Kewley \& Ellison 2008), 
we calculated the chemical composition of the HII regions C2 and C6 using 
multiple ratios of strong lines\footnote{
$R_{23}=\left(\left[OII\right]\lambda3727+\left[OIII\right]\lambda\lambda4959,5007\right)/H_{\beta}\lambda4861$, $O_{23}=\left(\left[OIII\right]\lambda4959+\left[OIII\right]\lambda5007\right)/\left[OII\right]\lambda3727$, $\left[NII\right]\lambda6584/H_{\alpha}\lambda6563$,
$\left[OIII\right]\lambda5007/\left[NII\right]\lambda6584$, $\left[NII\right]\lambda6584/\left[OII\right]\lambda3727$, and $\left[NII\right]\lambda6584/\left[SII\right]\lambda6720$}. 
For the R$_{23}$ method, the observed log$([NII]/H\alpha)>-1.1$ puts 
C2 and C6 in the upper branch. 

Then, following the procedure of Kewley \& Ellison (2008), we homogenized the derived values  
to the Pettini \& Pagel (2004) calibration to remove any systematic discrepancy between 
the various calibrations.
The abundances derived using several calibrations, before and after this 
conversion, are listed in Table \ref{Table7}. 
The inferred oxygen abundances show a peak to peak variation of $\sim 0.2-0.3$ dex.
By averaging these different values, we obtained our best estimate of the metallicity
$12+\log\left(\frac{O}{H}\right)=8.38$ for C2 and $12+\log\left(\frac{O}{H}\right)=8.29$ for C6.

Assuming a solar abundance $12+\log\left(\frac{O}{H}\right)=8.69$ 
(Asplund et al. 2009), these values imply a subsolar metallicity in these two 
regions ($Z=0.49Z_\odot$ for C2 and $Z=0.40Z_\odot$ for C6). 
Remarkably, this value is consistent with the metallicity derived for VCC1249 itself
using the mass metallicity relation (see subsection \ref{FitGal}),
which strengthens the hypothesis that the external HII regions were born in situ
from the pre-enriched gas that has been stripped from VCC1249.

\section{SED fitting}
\label{SED Fitting}

To constrain the star formation history of VCC1249 and its HII regions, 
we used the SED fitting technique presented in Fumagalli et al. (2011). 
First, using the spectrophotometric evolution code PEGASE2.0\footnote{
Projet d'Etude des GAlaxies par Synth\`ese Evolutive.}, 
we created a grid of synthetic spectra evaluated at different times for 
multiple input star formation histories SFR(t). Then,
using the SED-fitting code GOSSIP (Franzetti et al. 2008), 
we compared the observed photometric points (and spectra if available)
with the synthetic spectra. 
During this procedure, we analyzed VCC1249, M49, and the HII regions separately.

\subsection{The galaxy (VCC1249)} 
\label{FitGal}

As in Fumagalli et al. (2011), the evolution of the 
spectral energy distribution of the galaxy was modeled 
assuming a Salpeter IMF, with a lower mass end of $0.1$ M$_{\odot}$ and an  
upper mass of $120$ M$_{\odot}$. Initially, we assumed zero metallicity, 
then the ejecta of massive stars 
(Woosley \& Weaver 1995), implemented in the PEGASE code, enrich the ISM\footnote{PEGASE uses the 
``Padova'' stellar tracks, improved with the AGB phase 
(Groenewegen \& de Jong 1993) and with the helium white dwarfs 
(Althaus \& Benvenuto 1997). See 
http://www2.iap.fr/users/fioc/PEGASE.html for more details.}. 
Furthermore, we assumed a delayed exponential star formation history, dubbed 
``a la Sandage" (see equation 3 in Gavazzi et al. 2002, Sandage 1986),  
truncated as in Fumagalli et al. (2011) to simulate the stripping event on 
VCC1249 due to the interaction with M49. Analytically, this is
$$
\rm 
SFR(t) = \left\{ \begin{array}{rl}
 \frac{t}{\tau^2} e^{-\frac{t^2}{\tau^2}} &\mbox{ if $t<t_{trunc}$} \\
  0 &\mbox{ if $t>t_{trunc}$}
       \end{array} \right.
$$
where $t_{trunc}$ is the time from the onset until the end of star formation activity.

\begin{figure}[h!]
\begin{center}
\includegraphics[width=6.5cm]{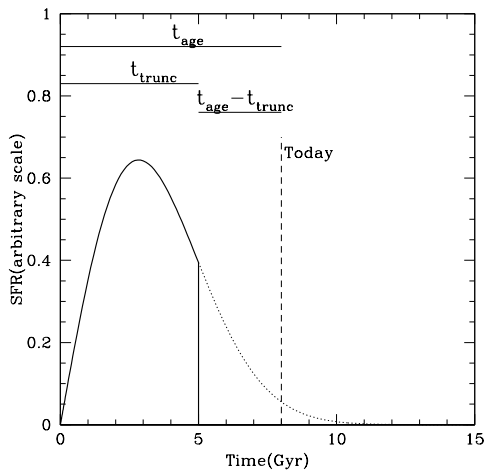} 
\caption{Example of star formation history from our library, with $\tau$=4 Gyr, $t_{age}$= 
8 Gyr and $t_{trunc}$ = 5 Gyr (adapted from Fumagalli et al. 2011).}
\label{Fig11}
\end{center}
\end{figure}

In Figure \ref{Fig11} (adapted from Fumagalli et al. 2011), we show an example 
of star formation history from our model library and we
highlight the various timescales that are relevant for this analysis.

In this grid of synthetic spectra, the characteristic time scale of the burst 
$\tau$ ranges from 1 to 20 Gyr with steps of 1 Gyr . The age of truncation $t_{trunc}$
varies from 1 to 13 Gyr with steps of 1 Gyr, while $t_{age}$ spans from 0 to 13.5 Gyr with steps of 100 Myr. 
Our final library includes 35K spectra. 
During the fit, we did not fix the age for the onset of star formation activity and, as previously
discussed, we did not include the (unknown) dust correction, since  
extinction is expected to be negligible for VCC1249 (see figure 8 
of Cortese et al. 2008 and 3-4 of Lee et al. 2009).

\begin{figure}
\begin{center}
\includegraphics[width=5.5cm]{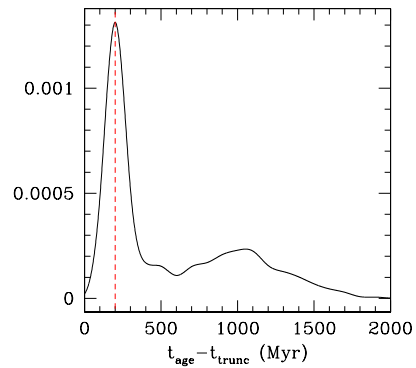} 
\caption{Probability distribution functions of the $t_{age}-t_{trunc}$ parameter for VCC1249. The 
truncation of star formation appears to be well constrained at about 200 Myr ago.
}
\label{Fig12}
\end{center}
\end{figure}

With this library, we ran the GOSSIP software to evaluate the parameters 
that best reproduce the observed photometry and their associated probability distribution functions (PDFs).
For VCC1249, we found that $t_{age}$, $t_{trunc}$, and $\tau$ are not well constrained individually
and we were only able to constrain the two latter parameters to be $<$ 2 Gyr. 
Conversely, the parameter $t_{age}-t_{trunc}$, which represents 
the lookback time at which the truncation of star formation occurred, is better constrained 
$t_{age}-t_{trunc} = 200 ^{+100} _{-100} \rm Myr$. Figure \ref{Fig12} shows the associated PDF.
This best-fit value agrees with the PSB spectral features 
observed in VCC1249 that are expected to arise  if the star 
formation activity halted abruptly in the past $0.5-1.5$ Gyr (Couch \& Sharples 1987).
Furthermore, this is consistent with the observed lack of H$\alpha$ emission and with the
presence of faint FUV emission.

\begin{table*}
\caption{
   Derived parameters of the HII regions. The SFR is computed from H$\alpha$ luminosity (left) and from FUV (right) for C1 and C2 and 
   only from FUV for the remaining regions using  Kennicutt (1998) calibrations. Mass and age are obtained from the SED fitting analysis. 
   }
\centering
\begin{tabular}{c|c|cc|cc}   
\hline
\hline
 Name   & Proj. Dist. from VCC1249 & Log H$\alpha$ Flux & SFR   & Mass    &  Age   \\
        &    (kpc)	  &   (erg  cm$^{-2}$ s$^{-1}$) & (10$^{-4}$ M$_{\odot}$yr$^{-1}$ )   & (10$^4$ M$_{\odot}$)     & 	(Myr)  \\
\hline 					   				     	  							    
       & 	  &	 	   &				    &	      & 		       \\  
C1     & 12.0     &  -15.06	   &  2.20 $\pm$ 0.09$/$7 $\pm$ 2	    &  1.04   &  4$^{+2}_{-1}$         \\ 	
       &	  &	 	   &				    &	      & 		       \\    	
C2     & 10.0     &  -14.83	   &  3.77 $\pm$ 0.11$/$6 $\pm$ 2	    &  1.45   &  7$^{+1}_{-1}$         \\	
       &	  &	 	   &				    &	      & 		       \\    	
C3     & 9.5      &  -  	   &  - 			    &  -      &   -			\\   	
       &	  &		   &	           		    &	      & 			\\   	
C4     & 11.0     &   - 	   &  1.7 $\pm$ 0.7         	    &  1.36   &  20$^{+30}_{-11}$	\\      
       &	  &		   &	           		    &	      & 			\\   	
C5     & 12.4     &   - 	   &  -            		    &  -      &  -			\\   	
       &	  &		   &	           		    &	      & 			\\   
C6     &  5.6     &    -	   &  6 $\pm$ 2         		    &  0.66   &  6$^{+1}_{-1}$  	\\   
       &	  &		   &	           		    &	      & 			\\   
C7     &  4.4     &   - 	   &  -            		    &  -      &  -			\\   
       &	  &		   &	           		    &	      & 			\\   
C8     &  4.9     &   - 	   &   7 $\pm$ 2        		    &  2.19   &  10$^{+3}_{-2}$ 	\\
       &	  &		   &	           		    &	      & 			\\   
C12    &   0.5    &  -  	   &  -            		    &  37.15  &  720$^{+520}_{-310}$	\\   
       &	  &		   &	           		    &	      & 			\\   
C17    &  1.5     &   - 	   &  -            		    &  17.78  &  2500$^{+980}_{-980}$	\\   
       &	  &		   &	           		    &	      & 			\\   
\hline
\end{tabular}
\
\label{Table8}
\end{table*}

Having inferred the star formation history of this galaxy, 
we can estimate its stellar mass by integrating the best-fit star formation history
$M_{star}=(1.20\pm0.16) \times 10^{9} \rm M_{\odot}$. This value is consistent with 
what was reported by Lee et al. 2003, $M_{star}=9.55 \times 10^{8} \rm M_{\odot}$.
For a 10$^9$ M$_\odot$  galaxy, the mass metallicity relation (Kewley \& Ellison 2008
using the Pettini \& Pagel 2004 calibration) predicts $12+\log\left(\frac{O}{H}\right)\sim 8.35$,
which agrees well with the metallicity derived for C2 and C6. Again, this confirms a scenario
in which the HII regions were born from the gas stripped from VCC1249 and pre-enriched 
by previous episodes of star formation.

\begin{figure}
\begin{center}
\includegraphics[width=9cm]{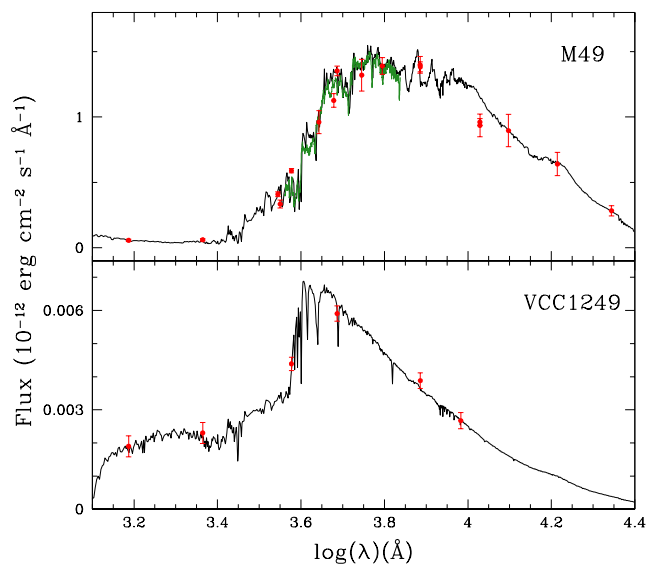} 
\caption{Model SED for M49 (top panel) together with the observed 
NIR photometric points and an optical spectrum (green) from GoldMine. 
The best-fit model is for a Sandage star formation history with $t_{age}=12$ Gyr and $\tau=2.8$ Gyr.
For VCC1249 instead (shown in the bottom panel), we find that a truncated Sandage star formation history
with a $t_{age}-t_{trunc}$ of  $\sim 200$ Myr provides a good description of the observed SED.}
\label{Fig15}
\end{center}
\end{figure}

Figure \ref{Fig15} shows the best-fit model and the SED of VCC1249 (bottom panel) and of 
M49 (top panel) 
( $\chi^2_{VCC1249}=0.29$; $\chi^2_{M49}=0.90$). 
A detailed analysis of the elliptical galaxy is 
beyond the scope of this study, but provides a test for our fitting tools. In this case, we complemented the NGVS 
measurements with J,H,K,U,B,V photometry and an optical spectrum taken from GoldMine. 
The best fit performed adopting a star formation history ``a la Sandage" and a Salpeter IMF provides 
$t_{age}=12^{+1}_{-2}$ Gyr and $\tau=2.8\pm1$ Gyr. 
Our determination agrees with the value 
of $t_{age}=11.9 $ Gyr found by Idiart et al. (2007) and marginally agrees 
with $t_{age}=9.6\pm1.4 $ Gyr of Thomas et al. (2005).

\begin{figure*}
\begin{center}
\includegraphics[height=14cm]{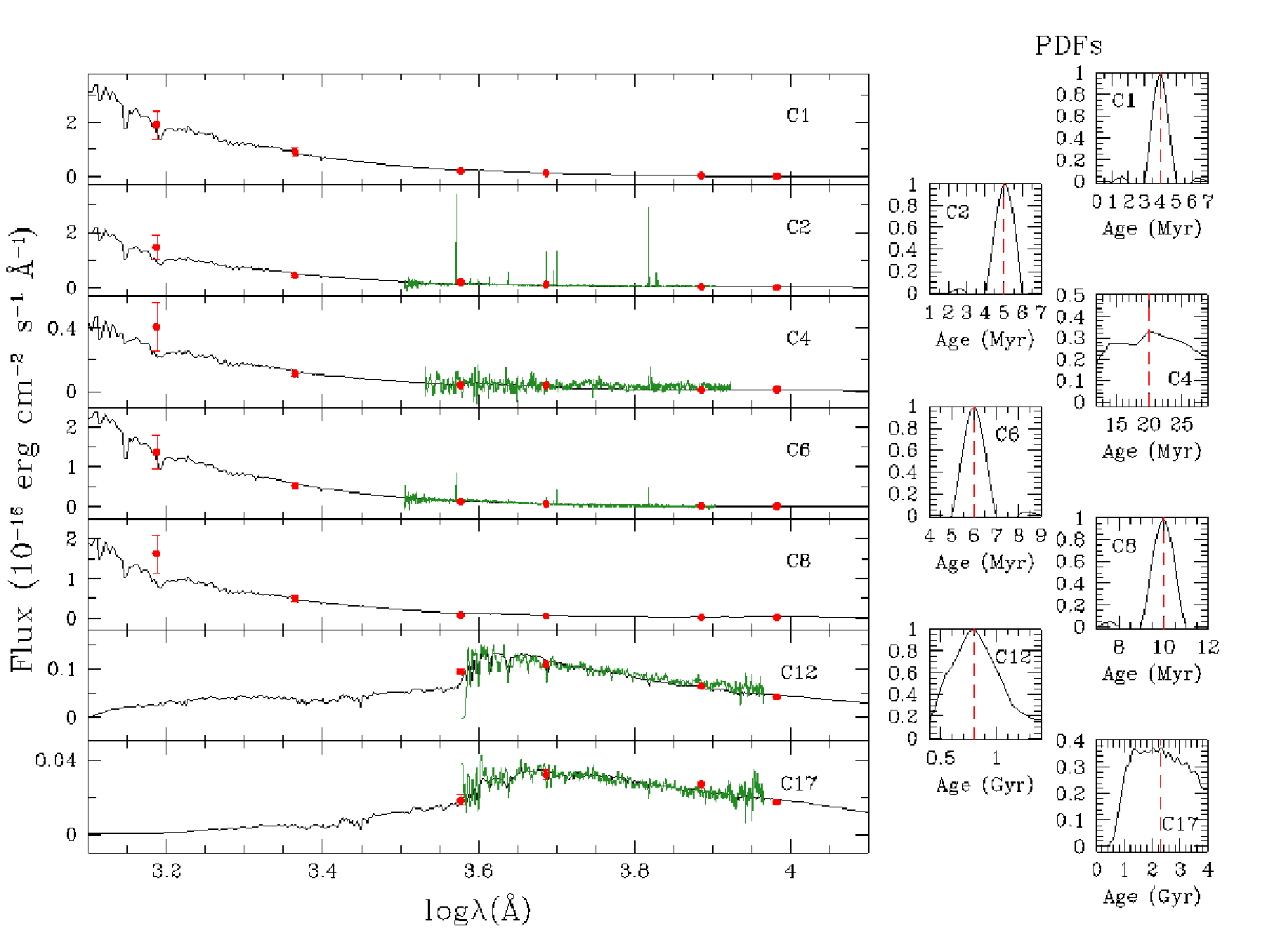} 
\caption{
Summary plot of the photometric points for the HII regions (red), observed optical spectrum if available 
(green) and their best-fit SEDs. Data for C1, C2, C4, C6, and C8 have been modeled using a single burst,
while C12 and C17 using an exponentially declining SFR. The probability distribution functions for the age 
parameter of each individual region are shown in the right panels. 
}
\label{Fig14}
\end{center}
\end{figure*}

\subsection{The star-forming regions}

For the analysis of the external HII regions we performed again a modeling 
similar to the one presented in Fumagalli et al. (2011). 
First, we computed a set of synthetic spectra with PEGASE2.0, assuming a single stellar burst 
with Salpeter IMF and slightly subsolar initial metallicity. As justified by the observed
Balmer decrement, we neglected internal dust absorption. 
For C12 and C17, which belong to VCC1249, we performed the SED fitting using both the single-burst model 
and an exponentially decreasing SFR with $\tau$ ranging from 10 to 2000 Myr.

Table \ref{Table8} contains the best-fit parameters for the observed SEDs. 
The outlying HII regions are well described (e.g.  $\chi^2_{C2}=0.63$;  $\chi^2_{C6}=0.24$) by 
a single burst of star formation, with age $\rm t < 30$ Myr. Instead, the 
star formation history of C12 and C17 is better described ($\chi^2_{C12}=1.47$;  
$\chi^2_{C17}=0.87$) by an exponential with $\tau<20$ Myr and with an age
about a factor of 100 higher than that of the HII regions 
outside the galaxy, such as C2 or C1.  
This result is consistent with the mean ages of $\approx 10^7$ yr 
and $10^8$ yr obtained by Lee et al. (1997) assuming the $\left(B-V\right)$-age relation given by 
Bruzual \& Charlot (1993).  

Figure \ref{Fig14} shows the SEDs of the regions, along with  
the PDFs of the age parameter that appears to be well constrained. 
As previously inferred from the lack of color variation among different HII regions 
(see Section \ref{MandPregions} and Figure \ref{Fig8}), 
our SED analysis indicates that these star-forming regions form an homogeneous 
population and that they are significantly different from VCC1249 and hence from 
the two knots belonging to the galaxy (C12 and C17).
As done previously for VCC1249, by integrating over the star formation history, we estimate 
the stellar masses of the studied regions to be between $0.66\times10^{4}$ and 
$3.72\times 10^{5}$ M$_{\odot}$. These values are typically found in star cluster complexes
and HII regions (Kennicutt et al. 1989).

\section{Discussion}
\label{Discussion}

Our multiwavelength analysis of VCC1249, based on new GALEX NUV data,
deep optical imaging (including  H$\alpha$) and Keck spectroscopy,
reveals that star formation was recently quenched in this dwarf  
and that the only star-forming regions are those outside the main body 
of the galaxy, made of pre-enriched gas in the HI cloud. 

These findings are consistent with the conclusions of 
Sancisi et al. (1987), Patterson \& Thuan (1992), and Lee et al. (2000) that
this dwarf has suffered from ram-pressure stripping in the hot atmosphere of M49, 
leading to the ablation of most of its original HI gas.
Following previous calculations (Sancisi et al. 1987; Patterson \& Thuan 1992) 
and using our newly estimated parameters, we find that VCC1249 is in fact unable to retain its gas 
at any radius because of ram pressure stripping.
The radius at which ram pressure (Gunn \& Gott 1972) becomes efficient can be 
estimated as (Domainko et al. 2006)
\begin{equation}
R_{strip} = 0.5 R_{0} ln \left (\frac{GM_{star}M_{gas}}{v^2 \rho_{Halo} 2 \pi R_{0}^4} \right),
\end{equation}
while the stripped mass is 
\begin{equation}
M_{strip} = M_{gas} \left(\frac{R_{strip}}{R_{0}}+1\right) exp\left(-\frac{R_{strip}}{R_{0}}\right).
\end{equation}
Here, $\rho_{Halo}= 10^{-3}$ cm$^{-3}$ is the gas density in the halo of M49 (Fabian 1985) and 
$v=611\pm32$ km s$^{-1}$ is the relative velocity between VCC1249 and M49.
In this calculation we furthermore assumed an exponential profile for the stellar and gas components, 
with $R_{0}=14.5$ arcsec, the inner scale length computed in Section \ref{Galaxy Morphology} using the 
$g$-band luminosity. Adopting $M_{star}=1.20 \times 10^9$ M$_{\odot}$, $M_{gas}= 8.41 \times 10^7$ M$_{\odot}$ 
(since $M_{gas}-M_{strip} = 1.51 \times 10^7$ M$_{\odot}$, Lee et al. 2003),
we find that VCC1249 cannot retain its gas at any radius, accordingly $M_{strip} =  M_{gas}$. 
Ram-pressure stripping 
can therefore fully deplete the gas reservoir of this dwarf, leading to a sudden
truncation of its star formation\footnote{We emphasize that the condition for the ram-pressure 
stripping is satisfied by using the relative velocity between the two galaxies along the line of sight. 
Possible motions in the plane of the sky would make ram pressure even more efficient.}.

However, as pointed out by Patterson \& Thuan (1992) and McNamara et al. (1994), ram-pressure 
stripping is not the only mechanism acting on this system because, for instance, 
it cannot explain the presence of the tail and counter-tail of VCC1249 (Fig. 2b,c).  
These features instead naturally arise from a tidal interaction with M49. 
Patterson \& Thuan (1992) estimated that the tidal radius, 
i.e. the distance at which stars in VCC1249 become unbound or stripped, is $>$ 20 arcsec, 
in agreement with the excess of light superimposed to the inner exponential profile 
that is visible in our deep optical profiles. We conclude therefore that 
both ram pressure and tidal interaction occurred during the interaction with M49: 
gravitational tides triggered the tail and the counter-tail of VCC1249 and aid ram pressure to 
remove the HI gas by diminishing the overall potential of the dwarf.

Furthermore, using our SED analysis, we can estimate the epoch of the encounter 
between VCC1249 and M49. Our best-fit model for the SED requires a sudden truncation 
of the star formation activity 200 Myr ago because of
gas ablation. This time is also consistent with an independent estimate for the 
ablation time derived by dividing the projected distance between the HI cloud and VCC1249 
(about 10 kpc) by their relative radial velocities ($\Delta V=79\pm30$ km s$^{-1}$).
Neglecting projection effects, this very crude calculation shows that 
$\sim 124\pm47$ Myr are needed to displace the gas from VCC1249 to the 
location where it is detected in the 21 cm emission.

Focusing on the outlying HII regions, both the SED fitting and the spectroscopy 
analysis reveal that their stellar populations are young  and coeval, being born within 
the last few tens of Myr. Furthermore, because all the outlying HII regions have an estimated age of 
less than 30 Myr, which is at least a factor of ten lower than 
the ablation time, we conclude that these star-forming regions
were born in situ after the removal of gas. This point is strengthened by the observed agreement between the oxygen abundance derived 
spectroscopically for the HII regions ($12+\log (O/H)=8.38$ for C2 and $12+\log (O/H)=8.29$ for C6) 
with the one for VCC1249 obtained using the mass-metallicity relation  ($12+\log (O/H)=8.35$).

The presence of a metal-enriched gas tail that extends to more than 3 arcmin far from 
VCC1249 in the direction of M49 is remarkable also in the context of the metal enrichment
of low-density gas. Indeed, within dense environments such as rich clusters, ram-pressure 
stripping may be an effective way to transport dust and metals from the ISM of galaxies to the 
outer intracluster medium (Boselli \& Gavazzi 2006; Cortese et al. 2010). Moreover, the encounter of VCC1249 with the halo of M49 may be 
the prototype of the interaction between satellite galaxies and their centrals and thus 
this mechanism may be responsible for at least part of the metal enrichment seen in the halos 
of galaxies at all redshifts (Tumlinson et al. 2011). This example shows that galactic winds may not be the only 
processes needed to transport metals to large galactocentric distances.

\subsection{Witnessing a peculiar interaction}

Witnessing this type of interaction in a cluster must be considered fortuitus, a fortiori
in the nearby Virgo cluster.
Considering the morphological types, systems like the VCC1249/M49 pair 
(a giant elliptical galaxy interacting with a dwarf irregular) are
rare in nearby clusters. In fact, giant elliptical (and cD) galaxies inhabit the center 
of clusters, while dwarf irregular galaxies tend to populate the cluster's outskirts.
The majority of dwarf galaxies suffer from one or more ram-pressure stripping events 
as they enter the cluster (Bekki 2009), leading to a complete 
removal of the atomic gas and to a consequent transformation into dwarf-ellipticals on very short time scales (Boselli et al. 2008a).
Thus, because of a combined effect of a lower velocity dispersion and gas density within the cluster, 
systems like VCC1249/M49 would be much more frequent at higher redshift, where 
the population of gas-rich dwarf galaxies has not yet been quenched by ram pressure.

However, looking at low redshift, some similarities with the studied system can be found in the famous 
interacting pair between the giant elliptical galaxy (M86) and a giant late-type galaxy (NGC 4438) in the 
Virgo cluster. The highly disturbed morphology of NGC 4438 is traditionally (e.g. Boselli et al. 2005) 
interpreted as caused by tidal interaction with its companion galaxy NGC 4435. 
Kenney et al. (2008), who discovered an extended complex of H$\alpha$ filaments connecting NGC 4438
with M86, instead invoked ram-pressure stripping and tidal (collision) interaction with M86. 
The ram pressure, resulting from the passage 
through the halo of M86, causes the removal of most of the ISM from the stellar disk of NGC 4438, 
while  tidal interaction produces most morphology disturbances in NGC 4438. 
As for VCC1249, NGC 4438 is very HI-deficient and what is believed to be the remnant of its atomic 
gas is observed near M86 (Li \& van Gorkom 2001). Both M86 and M49 have in their 
proximity an unusual cloud of HI, left-over of galaxies that passed through their halos.
However, nothing like the spectacular complex of H$\alpha$ filaments connecting NGC 4438 to M86, 
heated by thermal conductivity from the ICM and by turbulent shocks,
is present between VCC1249 and M49, where compact H$\alpha$ 
emission is detected  associated with star formation activity taking place in compact HII regions.
Thus, the presence of an HI cloud displaced in the halo of an elliptical galaxy, 
also combined with features visible in the UV-band and H$\alpha$, could be the 
signature of a rich-gas galaxy crossing.

\subsection{The complex diffuse structures around M49}

Additional information about the dynamical history of VCC1249 may be
gleaned from the tidal debris in the surrounding field. Early imaging
by Patterson \& Thuan (1992) and McNamara et al (1994) revealed a
debris trail $\sim$ 2\arcmin\ in length extending northwest from 
VCC1249, as well as a short countertail to the southwest. More recently,
Janowiecki et al (2010) used deep imaging to reveal a complex series
of accretion structures around M49, some possibly associated with 
VCC1249. With the deep NGVS imaging presented here, and the well-determined
photometric model for M49, we can now search even deeper to trace
these tidal features further in extent, and search for others.

To increase signal-to-noise at faint surface brightness and highlight
any diffuse structures found around M49 and VCC1249, we started with
our image of the field after subtracting the best-fit
elliptical isophotal model of M49 (shown in Figure \ref{lastFigure}). We then ran
IRAF's OBJMASKS task to identify and mask features that are
$10\sigma$ above a locally defined background measured on 1\arcmin\
scales. This effectively masks bright stars, the high surface
brightness regions of M49's companion galaxies, and a myriad of
fainter discrete sources on the image. We then spatially re-binned the
image in 39x39 (7.25\arcsec x7.25\arcsec) pixel bins, calculating the
median pixel intensity of each bin. The combination of masking and
medianing effectively rejects the contributions from small discrete
sources and maps the structure of diffuse light on $\sim$ 0.5 kpc
scales.

\begin{figure*}
\centering
\includegraphics[scale=0.68, angle=90] {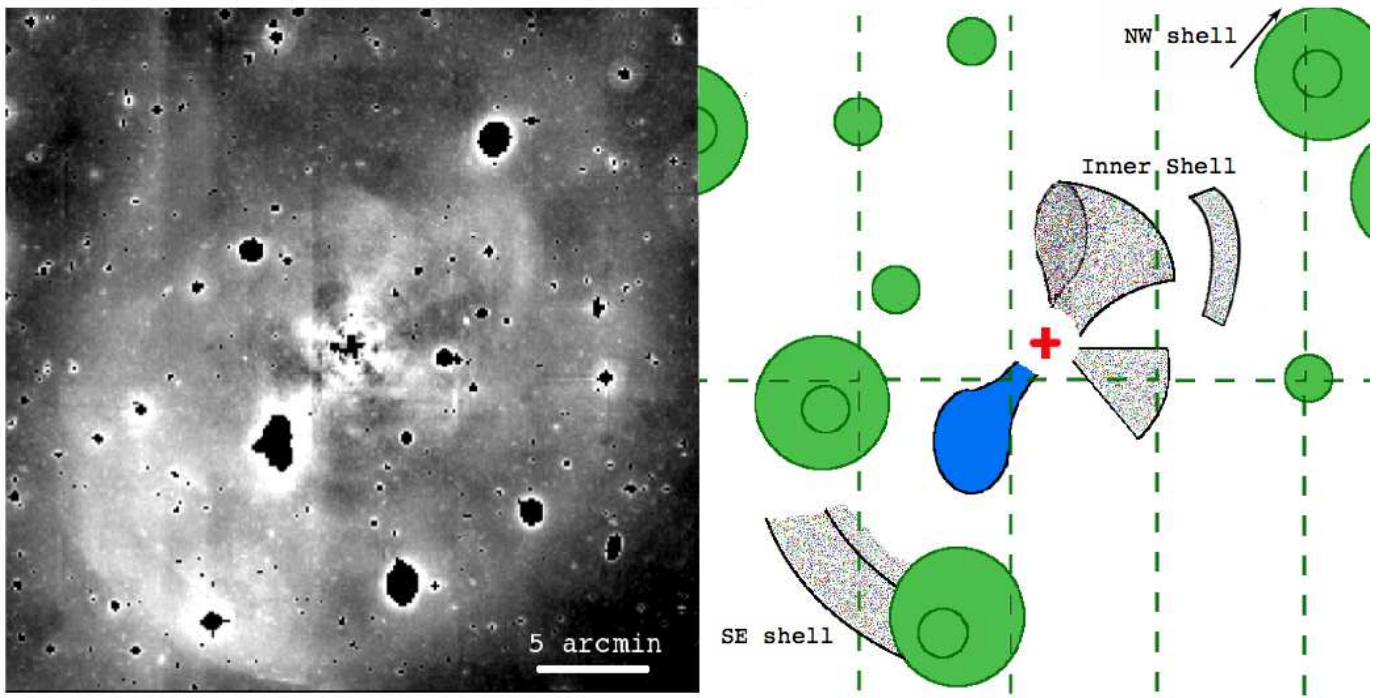} 
\caption{Left: binned, median-smoothed grayscale map of M49 showing residuals from the ELLIPSE model in the NGVS $g$-band
imaging. Although there are artifacts due to the edges of the CCDs and to the bright stars (circular reflection with two sizes: 
1 and 3 arcmin in radius), an extensive series of shells and filaments 
is apparent in agreement with Janowiecki et al. (2010). Right: toy-picture highlighting the diffuse features found in the left-image: 
the artifacts (green), shells and fan of material described in the text (gray) and VCC1249 with its tail pointing toward M49 (blue).  
}
\label{MihosMap}
\end{figure*}

Figure \ref{MihosMap} shows this binned, median-smoothed map. The vertical and
horizontal banding in the image are artifacts tracing the edges of the
CCDs on the Megacam array. In addition to these artifacts, several
diffuse features can be seen in the image. First, we trace a halo of
diffuse light around VCC1249 out to a major axis radius of $\sim$
2.4\arcmin\ (11 kpc). Beyond this radius, we see a stream of light
extending from VCC1249 to the northwest, toward the center of M49,
coincident with the extended HI and UV light. Around M49 itself, there
are a number of diffuse features; of particular interest are the
shells 5--10\arcmin\ to the northwest of M49, originally identified by
Janowiecki et al (2010; their ``inner shell'' feature) using deep
Schmidt imaging. Because the higher spatial resolution of Megacam
allows better rejection of background contaminants, these shells are
more clearly revealed in Figure \ref{MihosMap}, where they are resolved into two features. 
The innermost feature extending NW from the center of M49 shows a folded
fan-like shape, very reminiscent of features that arise from radial
accretion events (e.g., Hernquist \& Quinn 1988). A second shell can
be seen at slightly larger radius to the west-northwest of M49. Other
features can be seen in the image as well, including more extended
shells farther to the northwest and to the southeast, and a fan of
material extending west-southwest of the center of M49. Although 
the connection between these very extended NW and SE shells and VCC1249 is not 
clear (as Janowiecki et al. 2010 have already pointed out), 
the better resolution of our NGVS data shows much more detail in 
the ``inner shell'', and clearly links that feature to VCC1249. 
All these features are visible as well in the Schmidt imaging detailed in
Janowiecki et al (2010); the fact that they are visible in both
imaging surveys demonstrates that they are not simply artifacts generated by
instrumental effects such as reflections or scattered light.

We focus here only on the tidal features related to VCC1249 itself.
In this context, the folded fan-shaped plume to the northwest
of M49 is particularly intriguing. This type of structure forms from
the radial accretion of the disky companion, as detailed in Hernquist \&
Quinn (1988). Indeed, a comparison between the structure seen in
Figure \ref{MihosMap} and the ``t=225'' snapshot of the HQ88 disk accretion model
(their Figure 2) is striking. In this scenario, VCC1249 has made (at
least) two close passages past M49 with the radius of the inner NW
shell showing the apocentric turning point from the previous
passage. This shell occurs at a projected radius of 6.4\arcmin\, or 30
kpc from M49's center. At this radius, the M49 mass model of C{\^o}t{\'e} et
al. (2003) has an enclosed mass of $2\times 10^{12}$ M\sun, yielding a
dynamical time of $\sim$ 100 Myr. Since VCC1249 is currently found projected
5.6\arcmin\ (26 kpc) southeast of M49, this is also a rough lower
limit on the time since the last closest passage to M49. Of course, this
limit assumes that VCC1249's orbit lies along the plane of the
sky and the projected separations are true separations, which is
likely not true. The fact that M49 and VCC1249 have a velocity
difference of 611 km s$^-1$ argues that much of the orbital motion is along
the line of sight, which significantly increases the inferred time since
last passage. Given these uncertainties, a scenario where VCC1249 is
on a fairly radial, bound orbit with last pericenter passage occurring
a few hundred Myr ago is quite reasonable; this orbital timescale is
quite similar to the star formation truncation time we derived earlier (see Section \ref{FitGal})
and argues for a common origin\footnote{The timescale since the first passage is much longer 
than the  derived truncation time. However, the SED analysis can only detect the last truncation 
event, the second passage.}. In this scenario, tidal stripping of
material from VCC1249 is ongoing, as is ram-pressure stripping. Indeed, because according to 
this scenario the dwarf galaxy is on its second passage through the dense halo of M49, 
it likely lost a significant amount of gas on the first passage as well. This first 
passage could have stripped the outer, low-density gas in VCC1249, leaving the denser gas 
in the inner regions to be stripped on this subsequent passage we observe now.

\subsection{Formation of compact objects}

In addition to the interest for studies of galaxy interactions in rich clusters, 
isolated HII regions born from ram-pressure stripped gas provide some insight into 
the formation of evolved compact systems such as globular clusters (GC) and ultra compact dwarf galaxies (UCD).
Found preferentially in the neighbors of massive elliptical galaxies (Hilker 2011), the origin of the latter 
compact systems is still controversial. Models indicate that they can form either 
by tidal stripping of the diffuse, low surface brightness disks of nucleated dwarf ellipticals
(Bekki et al. 2003), or might derive from the merging of many 
young, massive (10$^6$ M$_\odot$) star clusters, born during a past merger event 
(Kroupa 1998; Fellhauer \& Kroupa 2002; Kissler-Patig et al. 2006).
The HII regions formed during the interaction of the (originally) star-forming VCC1249 with 
M49 might be the progenitors of these compact systems.
They have very compact morphologies and are composed of coeval stars formed during 
a single episode of star formation. 
The stripped gas that gave birth to these HII regions is now (dynamically) 
dissociated from them and thus it cannot sustain any more star formation. Given their mass 
($\sim$ 10$^4$ M$_\odot$) and their young age ($\simeq$ 5-10 Myr), it is still unclear whether 
they will 
survive the "infant mortality" caused by the kinetic energy injected by supernova explosions (Lada \& Lada 2003).
Recent models seem to indicate that at least a fraction of them should resist 
and end up in compact evolved systems (Gieles \& Bastien 2008).
Containing only 10$^4$ M$_\odot$, the HII regions found around M49 certainly do not have the necessary mass to 
form UCDs or massive GCs, but only intermediate-mass compact objects. Considering however that at high redshift 
late-type galaxies had gas masses significantly higher than those observed in the local universe 
(e.g. Boselli et al. 2001), the HII
regions formed during the interaction of VCC1249 with M49 
could be the scaled-down version at $z=0$ of more massive star clusters formed at 
\begin{figure*}
\centering
\includegraphics[scale=0.925]{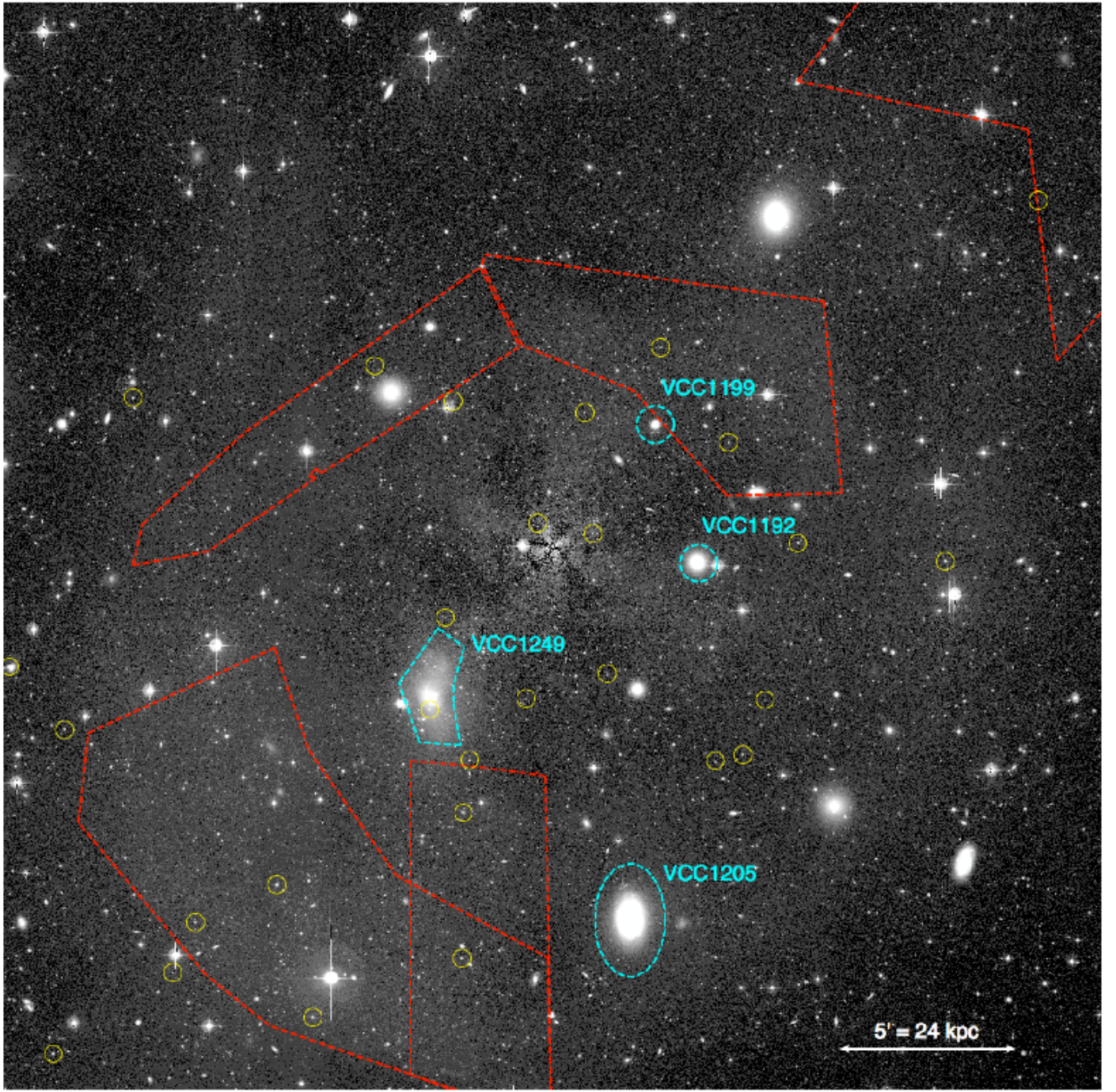} 
\caption{Grayscale image of M49 showing residuals from the ELLIPSE model that best fits the azimuthally averaged isophotes in the NGVS $g$-band
imaging (see Ferrarese et al. 2011).  An extensive series of shells and filaments is apparent. A complex structural was also found by 
Janowiecki et al. (2010); the dashed red lines indicate the regions where these authors found shells and plumes in their residual image 
(here shown for comparison with those in Figure \ref{MihosMap}). 
VCC1249 is labeled in cyan, as are VCC1199 and VCC1192, two compact elliptical galaxies that have likely undergone tidal stripping
(e.g., Cote et al. 2010). VCC1205 shows evidence for star formation detached from the main body of the galaxy, in the direction of M49.
Yellow circles show the position of candidate UCDs (having $g \le 21$ and effective radii in the range $10 \le R_e \le 100$~pc) identified
from the NGVS imaging. At least some of these objects may have formed through tidal ``threshing" of nucleated dwarf galaxies (e.g.,
Bekki et~al. 2001).
}
\label{lastFigure}
\end{figure*}
higher redshift that later transformed into today massive GCs and UCDs. Similar star-forming compact structures, 
but of significantly higher mass, 
have been indeed observed in the tails of some massive spirals in the clusters 
A1689 and A2667 at $z$ $\simeq$ 0.2 by Cortese et al. (2007).

\subsection{Conclusion}

In conclusion, the interaction between VCC1249 and M49 underscores the fundamental role played by environment in shaping the properties of
galaxies in cluster cores. In this particular case, VCC1249 is undergoing both ram pressure and tidal interaction: the joint action of 
the two mechanisms leads to the removal of the HI gas, while the morphology disturbances are triggered by the gravitational tides. 
Our analysis suggests that the star formation in VCC1249 was truncated 200 Myr ago, which is consistent with the gas ablation time. 
The HII regions were born {\it in situ}, within the turbulent, pre-enriched gas that was removed by the interaction.

\subsection{A harsh environment: the Virgo B subcluster}

\begin{figure}
\begin{center}
\includegraphics[scale=0.4]{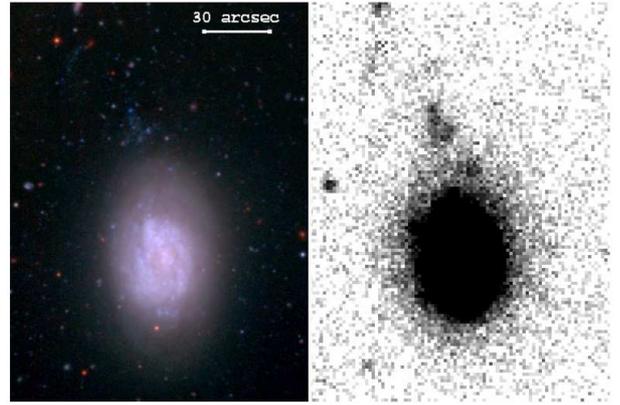} 
\caption{Left: RGB image of VCC1205 obtained by combining the NGVS images 
in the $u, g, z$ filters. Right: NUV image of VCC1205 on the same scale. Blue extended 
star-forming regions are visible in the north direction (toward M49).}
\label{theBest}
\end{center}
\end{figure}

We conclude with a panoramic view of the immediate neighborhood of VCC1249 --- an environment that is, of course, dominated by M49, the 
brightest member of the Virgo cluster and the central galaxy in Virgo's B subcluster (e.g., Binggeli, Popescu \& Tammann 1993).
Our analysis of VCC1249 adds to the growing body of evidence that interactions, mergers and stripping have had a profound affect on
the galaxy population in this high-density environment. Figure \ref{lastFigure} illustrates some of this evidence. In addition to VCC1249, this 
region also contains two examples of the rare class of ``compact elliptical" galaxies, whose origin is almost certainly related to strong
tidal stripping of initially more massive galaxies (e.g., Faber 1973; Bekki et~al. 2001b; Chilingarian et al. 2009; C\^ot\'e 2010;
Huxor et al. 2011). Furthermore, as shown in the residual image in Figures \ref{MihosMap} and \ref{lastFigure}, a complex series of shells, 
plumes and streams surrounds 
M49 --- {\it prima facie} evidence for past accretions and interactions (Janiowiecki et al. 2010; Ferrarese et al. 2011). Fitting of PSF-convolved
models to sources in the NGVS images also reveals a large number of UCDs candidates, which are shown as the yellow circles in Figure 
\ref{lastFigure} (see also Ha{\c s}egan et al. 2005).
Although there may well be multiple formation channels for UCDs, at least some of these objects could have formed through tidal ``threshing"
of nucleated dwarf galaxies, a leading UCD formation mechanism (Bekki et al. 2001a). Finally, approximately 11$^{\prime}$ to the south
of M49 lies VCC1205, which is classified as ScIII-pec by Binggeli et al. (1985). Despite its high relative velocity with respect to M49
($\Delta v = v_{\rm M49} - v_{\rm VCC1205} = 1001 - 2341 = 1340~$km s$^-1$),
the galaxy shows a minor HI deficiency, while its optical and NUV morphology shows evidence for star-forming regions
detached from the main body of the galaxy in the direction of M49 (Figure \ref{theBest}). This is perhaps another indication of the harsh environment existing
in the core of the Virgo subcluster B. In any case, there is mounting evidence that the central regions of rich clusters --- and of the
Virgo B subcluster in particular --- are highly dynamic environments in which interactions influence the structure of both the central
galaxies and their surrounding satellites.

\section{Acknowledgments}

We warmly thank Mattia Fumagalli for his precious contribution on the SED fitting procedures.
We really appreciated the help provided by J. Xavier Prochaska in obtaining Keck data. 
This work made extensive use of GoldMine, 
the Galaxy On Line Database (http://goldmine.mib.infn.it).
We are grateful to P. Franzetti and M. Hilker for constructive discussions. 
We thank Joseph F. Hennawi for his useful comments on the draft.\\
Some of the data presented herein were obtained at the W.M. Keck Observatory, 
which is operated as a scientific partnership among the California Institute of Technology, 
the University of California and the National Aeronautics and Space Administration. 
The Observatory was made possible by the generous financial support of the W.M. Keck Foundation.
The authors wish to recognize and acknowledge the very significant cultural role and reverence 
that the summit of Mauna Kea has always had within the indigenous Hawaiian community. 
We are most fortunate to have the opportunity to conduct observations from this mountain.
The present study could not have been conceived without the DR7 of SDSS. 
Funding for the Sloan Digital Sky Survey (SDSS) and SDSS-II has been provided by the 
Alfred P. Sloan Foundation, the Participating Institutions, the National Science Foundation, 		    
the U.S. Department of Energy, the National Aeronautics and Space Administration, 			    
the Japanese Monbukagakusho, and 									    
the Max Planck Society, and the Higher Education Funding Council for England. 				    
The SDSS Web site is http://www.sdss.org/.								    
The SDSS is managed by the Astrophysical Research Consortium (ARC) for the Participating Institutions.      
The Participating Institutions are the American Museum of Natural History, Astrophysical Institute Potsdam,  
University of Basel, University of Cambridge, Case Western Reserve University, The University of Chicago,   
Drexel University, Fermilab, the Institute for Advanced Study, the Japan Participation Group, 		    
The Johns Hopkins University, the Joint Institute for Nuclear Astrophysics, the Kavli Institute for 	    
Particle Astrophysics and Cosmology, the Korean Scientist Group, the Chinese Academy of Sciences (LAMOST),  
Los Alamos National Laboratory, the Max-Planck-Institute for Astronomy (MPIA), the Max-Planck-Institute     
for Astrophysics (MPA), New Mexico State University, Ohio State University, University of Pittsburgh, 	    
University of Portsmouth, Princeton University, the United States Naval Observatory, and the University     
of Washington.\\											    
GALEX is a NASA Small Explorer, launched in 2003 April. We gratefully 
acknowledge NASA's support for construction, operation and science 
analysis for the GALEX mission, developed in cooperation with the 
Centre National d'Etudes Spatiales (CNES) of France and the Korean 
Ministry of Science and Technology.\\
The research leading to these results has received funding from the 
European Community's Seventh Framework Programme (/FP7/2007-2013/) under 
grant agreement No 229517.\\
This work is supported in part by the Canadian Advanced Network for 
Astronomical Research (CANFAR) 
which has been made possible by funding from CANARIE under the 
Network-Enabled Platforms program.\\
G. Gavazzi acknowledges financial support from Italian MIUR PRIN contract 200854ECE5 
and from the high energy contract ASI-INAF I/009/10/0. 
J.C. Mihos thanks the National Science Foundation for support through 
awards ASTR-0607526 and AST-0707793.\\
We thank the Referee for the thorough reading of the manuscript and helpful comments.

\clearpage

\end{document}